 \documentclass[12pt]{article}
 \usepackage{graphicx,amsmath,latexsym,amsfonts}

 \font\ro=cmsy10                          
 \def\kcr{{\hbox{\ro \char'170}}}                
 \def\ktl{{\hbox{\ro \char'170}}}        
 \def\ktr{{\hbox{\ro \char'170}}}        
 \def\kbl{{\hbox{\ro \char'170}}}        
 \def\kbr{{\hbox{\ro \char'170}}}        

 \parskip=\medskipamount                 
 \thicklines                         

 \def\hat{\widehat}
 \def\Z{\mathbb{Z}}

 \def\R{\mathbb{R}}

 \def\bpl{\Big(}
 \def\bpr{\Big)}
 \def\e{\epsilon}
 \def\ve{\varepsilon}
 \def\t{\theta}

 \def\der{\partial}
 \newcommand{\ft}[2]{{\textstyle\frac{#1}{#2}}}
 \def\brr{\begin{equation}}
 \def\err{\end{equation}}
 \def\brr{\begin{eqnarray}}
 \def\err{\end{eqnarray}}
 \def\ba{\left(\begin{array}}
 \def\ea{\end{array}\right)}
 \def\lf{\left.\begin{array}{c}}
 \def\rf{\end{array}\right.}

 \newcommand{\dr}{\raise.3ex\hbox{$\stackrel{\leftarrow}{\partial }$}{}}
 \newcommand{\dl}{\raise.3ex\hbox{$\stackrel{\rightarrow}{\partial}$}{}}
 \newcommand{\topi}{\raise.3ex\hbox{$\stackrel{\pi}{\longrightarrow}$}{}}
 
 \renewcommand{\theequation}{\arabic{section}.\arabic{equation}}
 \renewcommand{\a}{\alpha}

 \renewcommand{\d}{\delta}
 \newcommand{\s}{\sigma}

 \def\oldheadpic{                                
        \setlength{\unitlength}{.4mm}
        \thinlines
        \par
        \begin{picture}(349,16)
        \put(325,16){\line(1,0){4}}
        \put(330,16){\line(1,0){4}}
        \put(340,16){\line(1,0){4}}
        \put(335,0){\line(1,0){4}}
        \put(340,0){\line(1,0){4}}
        \put(345,0){\line(1,0){4}}
        \put(329,0){\line(0,1){16}}
        \put(330,0){\line(0,1){16}}
        \put(339,0){\line(0,1){16}}
        \put(340,0){\line(0,1){16}}
        \put(344,0){\line(0,1){16}}
        \put(345,0){\line(0,1){16}}
        \put(329,16){\oval(8,32)[bl]}
        \put(330,16){\oval(8,32)[br]}
        \put(339,0){\oval(8,32)[tl]}
        \put(345,0){\oval(8,32)[tr]}
        \end{picture}
        \par
        \thicklines
        \vskip.2in}
 \def\oldtitle#1#2#3#4{\oldheadpic\begin{center}\vglue.5in{\large\bf #1}\\[.6in]
        {#2}\\[.1in] {\it Department of Physics and Astronomy}\\
        {\it University of Maryland, College Park, MD 20742}\\[.6in]
        Physics Publication \#{#3}\\ {#4}\\[1.5in] {\bf ABSTRACT}\\[.1in]
        \end{center} \begin{quotation}}                 
 \def\oldTitle#1#2#3#4#5#6#7{\oldheadpic\begin{center} \vglue .4in
        {\large\bf #1}\\[.4in]
        {#2}\\[.1in] {\it Department of Physics and Astronomy}\\
        {\it University of Maryland, College Park, MD 20742}\\[.1in]
        {#3}\\[.1in] {\it {#4}}\\ {\it {#5}}\\[.4in]
        Physics Publication \#{#6}\\ {#7}\\[.5in] {\bf ABSTRACT}\\[.1in]
        \end{center} \begin{quotation}}                 
 \def\border{                                            
        \setlength{\unitlength}{1mm}
        \newcount\xco
        \newcount\yco
        \xco=-21
        \yco=12
        \begin{picture}(140,0)
        \put(\xco,\yco){$\ktl$}
        \advance\yco by-1
        {\loop
        \put(\xco,\yco){$\kcr$}
        \advance\yco by-2
        \ifnum\yco>-240
        \repeat
        \put(\xco,\yco){$\kbl$}}
        \xco=158
        \yco=12
        \put(\xco,\yco){$\ktr$}
        \advance\yco by-1
        {\loop
        \put(\xco,\yco){$\kcr$}
        \advance\yco by-2
        \ifnum\yco>-240
        \repeat
        \put(\xco,\yco){$\kbr$}}
        \put(-20,13){\tiny **University of Maryland * Center for String and
        Particle  Theory* Physics Department***University of Maryland*Center
        for String and Particle Theory** }
        \put(-20,-241.5){\tiny ***
        Physics Department ***
        Hobart and William Smith Colleges ** Physics Department **
        Hobart and William Smith Colleges *** Physics Department *** }
        \end{picture}
        \par\vskip-8mm}
 \def\bordero{                                           
        \setlength{\unitlength}{1mm}
        \newcount\xco
        \newcount\yco
        \xco=-31
        \yco=12
        \begin{picture}(140,0)
        \put(\xco,\yco){$\ktl$}
        \advance\yco by-1
        {\loop
        \put(\xco,\yco){$\kclr}
        \advance\yco by-2
        \ifnum\yco>-240
        \repeat
        \put(\xco,\yco){$\kbl$}}
        \xco=151
        \yco=12
        \put(\xco,\yco){$\ktr$}
        \advance\yco by-1
        {\loop
        \put(\xco,\yco){$\kcr$}
        \advance\yco by-2
        \ifnum\yco>-240
        \repeat
        \put(\xco,\yco){$\kbr$}}
        \put(-20,12){\ooo bacdefghidfghghdhededbihdgdfdfhhdheidhdhebaaahjhhdahbahgdedgehgfdiehhgdigicba}
        \put(-20,-241.5){\oooababaighefdbfghgeahgdfgafagihdidihiidhiagfedhadbfdecdcdfagdcbhaddhbgfchbgfdacfediacbabab}
        \end{picture}
        \par\vskip-8mm}
 \def\headpic{                                           
        \indent
        \setlength{\unitlength}{.4mm}
        \thinlines
        \par
        \begin{picture}(29,16)
        \put(165,16){\line(1,0){4}}
        \put(170,16){\line(1,0){4}}
        \put(180,16){\line(1,0){4}}
        \put(175,0){\line(1,0){4}}
        \put(180,0){\line(1,0){4}}
        \put(185,0){\line(1,0){4}}
        \put(169,0){\line(0,1){16}}
        \put(170,0){\line(0,1){16}}
        \put(179,0){\line(0,1){16}}
        \put(180,0){\line(0,1){16}}
        \put(184,0){\line(0,1){16}}
        \put(185,0){\line(0,1){16}}
        \put(169,16){\oval(8,32)[bl]}
        \put(170,16){\oval(8,32)[br]}
        \put(179,0){\oval(8,32)[tl]}
        \put(185,0){\oval(8,32)[tr]}
        \end{picture}
        \par\vskip-6.5mm
        \thicklines}
 \def\title#1#2#3#4{\border\headpic {\hbox to\hsize{#4 \hfill UMDEPP #3}}\par
        \begin{center} \vglue .5in {\large\bf #1}\\[.6in]
        {#2}\\[.1in] {\it Department of Physics and Astronomy}\\
        {\it University of Maryland, College Park, MD 20742}\\[1.5in]
        {\bf ABSTRACT}\\[.1in] \end{center} \begin{quotation}}  
 \def\Title#1#2#3#4#5#6#7{\border\headpic
        {\hbox to\hsize{#7 \hfill UMDEPP #6}}\par
        \begin{center} \vglue .4in {\large\bf #1}\\[.4in]
        {#2}\\[.1in] {\it Department of Physics and Astronomy}\\
        {\it University of Maryland, College Park, MD 20742}\\[.1in]
        {#3}\\[.1in] {\it {#4}}\\ {\it {#5}}\\[.5in] {\bf ABSTRACT}\\[.1in]
        \end{center} \begin{quotation}}                 
 \def\endtitle{\end{quotation}\newpage}                  

 \def\qd{{\kern0.5pt
                   q \kern-5.05pt \raise5.8pt\hbox{$\textstyle.$}\kern 0.5pt}}

\begin{document}
 \border\headpic {\hbox to\hsize{July 2004 \hfill {HWS-2004A07}}}
 \par
 {$~$ \hfill {UMDEPP 05-007}}
 \par
 {$~$ \hfill {hep-th/yymmnn}}
 \par

 \setlength{\oddsidemargin}{0.3in}
 \setlength{\evensidemargin}{-0.3in}
 \begin{center}
 \vglue .05in {\large\bf Adinkras: A Graphical Technology for\\[.1in]
 Supersymmetric Representation Theory\footnote
 {Supported in part  by National Science Foundation Grant PHY-0354401.}\  }
 \\[.25in]
 Michael Faux\footnote{faux@hws.edu}
 \\[0.06in]

 {\it Department of Physics\\
 Hobart and William Smith College\\
 Geneva, NY 14456, USA}\\[.03in]
 and
 \\[.05in]
 S.\, J.\, Gates, Jr\footnote{gatess@wam.umd.edu}
 \\[0.06in]

 {\it Center for String and Particle Theory\\
 Department of Physics, University of Maryland\\
 College Park, MD 20742-4111 USA}\\[1.5in]

 {\bf ABSTRACT}\\[.01in]
 \end{center}
 \begin{quotation}
 {We present a symbolic method for organizing the representation theory
 of one-dimensional superalgebras.  This relies on special objects, which we
 have called adinkra symbols, which supply tangible geometric
 forms to the still-emerging mathematical basis underlying supersymmetry.}

 ${~~~}$ \newline PACS: 04.65.+e

 \endtitle

 \section{Introduction}

 ~~~~ There are important examples in which theoretical physics
 incorporates elegant motifs to represent mathematical conceptions
 that are vastly simplified thereby.  One such example is
 the wide-spread use of Feynman diagrams.  Another one of these is
 Salam-Strathdee superspace,
 a stalwart construction which has proven most helpful
 in organizing fundamental notions in field theory and in string theory.
 Despite its successes, however, there are vexing
 limitations which bedevil attempts to use this latter construction to
 understand certain yet-mysterious aspects of off-shell supersymmetry.
 This situation would seemingly benefit from an improved organizational scheme.
 In this paper, we introduce a graphical paradigm which
 shows some promise in providing a new symbolic technology for
 usefully re-conceptualizing problems in supersymmetric representation
 theory.

 The use of symbols to connote ideas which defy
 simple verbalization is perhaps one of the oldest of human
 traditions. The Asante people of West Africa have long been
 accustomed to using simple yet elegant motifs known as Adinkra
 symbols, to serve just this purpose.  With a nod to this
 tradition, we christen our graphical symbols as ``Adinkras.''

 Our focus in this paper pertains most superficially to the
 classification of off-shell representations of arbitrary
 $N$-extended one-dimensional superalgebras.  However, for
 some time, we have been aware of evidence that suggests that
 {\it every} superalgebra, in any spacetime dimension, has its
 representation theory fully encoded in the representations of
 corresponding one-dimensional superalgebras.  This idea, and
 much of the relevant mathematical technology for substantiating
 this idea, has been developed in a series of previous papers
 \cite{GatesLP, GatesRana1, GatesRana2}.  One purpose of this current
 paper is to introduce a new notational tool which, among other things,
 adds tangible conceptual forms useful for discerning both the content
 and the ramifications of this mathematics.  The tool we introduce is
 a new sort of symbol which usefully represents supermultiplets.

 The relevance of our investigation extends beyond the realm of
 representation theory, however.  Indeed, there are reasons to suppose that
 supersymmetric quantum mechanics might include undiscovered algebraic
 structures related to interesting fundamental questions.  Consider
 the simple observation that every quantum field theory formulated in
 any spacetime dimension, has a corresponding supersymmetric quantum mechanical
 model obtained by dimensionally reducing all of the spatial dimensions.
 We refer to these quantum mechanical models as ``shadows"
 of the original quantum field theories.
 \begin{figure}
 \begin{center}
 \includegraphics[width=3.5in]{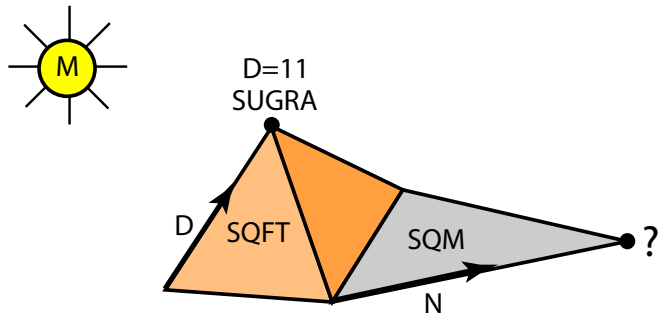}
 \caption{Each supersymmetric quantum field theory has a ``shadow" in
 supersymmetric quantum mechanics obtained by dimensionally reducing
 all of the spatial dimensions in the field theory.}
 \label{pyramid}
 \end{center}
 \end{figure}
 Higher spacetime dimension $D$ manifests in the shadow
 version as higher $N$, whereas structure group $SO(D-1,1)$
 transformations manifest as $R$-transformations.
 Those quantum field theories having remarkable algebraic features,
 anomaly cancellation for example, must have algebraically-interesting
 shadows as well.
 Since eleven-dimensional supergravity is a unique theory,
 the corresponding $N=32$ supersymmetric quantum mechanics certainly
 exhibits its own special uniqueness.
 One might wonder how the
 feature of anomaly freedom in effective string theory
 descriptions of ten-dimensional supergravity
 manifest on corresponding shadow mechanics.
 This viewpoint might be useful in discerning whatever
 analogs of string-theory modular invariance exist in
 {\it M}-theory.

 We should emphasize the importance of finding an overarching
 off-shell representation theory for supersymmetry.  This is a problem
 that has been largely ignored as theoretical physicists have been
 able to uncover ever more interesting and complicated theories that
 involve supersymmetry by ever more creative means.  We refer to
 this as the ``auxiliary field problem.''  Some familiar
 systems in which this problem is observed are 11D supergravity
 and all known 10D supersymmetric systems.  Since each of these particular
 systems are special limits of closely related {\it M}-theory and 10D
 superstrings, it follows that any increase in our understanding of these
 special limits is likely to accrue benefits to our understanding of the
 full theories.

 This paper is structured as follows.

 In sections 2 through 7 we present an overview of the mathematical
 basis for the core part of the paper, which begins in section 8.  The
 review sections are included in part to make this paper relatively
 self-contained.  But these also include several important new definitions
 and include commentary which may prove helpful to the reader.  In
 these sections we describe the elemental superalgebra and set our notational conventions.
 We review that class of irreducible representations which includes generalized scalar and generalized
 spinor multiplets, and discuss aspects of automorphic duality transformations.
 We review the connection between the multiplets mentioned above and the algebras ${\cal GR}(d,N)$.
 We review the connection between the representations of ${\cal GR}(d,N)$
 and those of the Clifford algebras $C(N,1)$.  We then review the notion
 of a root superfield proposed in \cite{GatesLP} which may provide the mathematical
 lynch pin for the classification of all supermultiplets.

 In sections 8 through 14 we methodically develop the conception
 of adinkra symbols referred to above. In successive sections, we
 show how elemental $N=1$ adinkra symbols can be combined to
 describe higher-$N$ representations, and how duality maps connect
 these with adinkras describing distinct multiplets.  We show how the adinkra symbols
 fit naturally into the concept of a root space and how
 supersymmetry transformations can be viewed in terms of flows
 on a lattice.  We use these techniques to describe new multiplets
 which exhibit interesting topological distinctions.  We use these techniques to
 comprehensively describe all of the known irreducible
 multiplets for $N\le 4$, and a few interesting reducible multiplets.  In
 so doing, we are hopeful that the discussion presents a satisfying
 re-conceptualization of traditional superspace reduction techniques,
 and a satisfying re-conceptualization of gauge invariance in
 supermultiplets.

 \setcounter{equation}{0}
 \section{The Elemental $d=1$ Superalgebra}

~~~~ The most basic of all superalgebras is the $d=1$ $N=1$
 superalgebra, which can be written as
 \brr [\,\d_Q(\e_1)\,,\,\d_Q(\e_2)\,]=
      -2\,i\,\e_1\,\e_2\,\der_{\tau} \,,
 \label{basic12}\err
 where $\e_{1,2}$ are real anticommuting parameters and $\tau$
 is a ``proper time" which parameterizes the one-dimensional space.
 We like to interpret this space as the worldline traced out by a particle
 in an ambient ``target-space".   There are two irreducible representations
 of (\ref{basic12}).  The first of these is the $d=1$ $N=1$ scalar multiplet,
 which includes a real commuting field $\phi$ ($\phi$ = $\phi^*$ where
 the $*$-operation denotes ``superspace conjugation) as lowest component and a
 real anticommuting field $\psi$ ($\psi$ = $\psi^*$) as highest component.
 The other basic multiplet is the $d=1$ $N=1$ spinor multiplet, which includes a
 real anticommuting field $\eta$  as a lowest component and a real commuting field $B$ as highest component.
 The supersymmetry transformation rules  are
 \brr \d_Q\,\phi &=& i\,\e\,\psi
      \hspace{1in}
      \d_Q\,\eta =
      \e\,B
      \nonumber\\[.1in]
      \d_Q\,\psi &=&
      \e\,\dot{\phi}
      \hspace{1in}
      \d_Q\,B = i\,\e\,\dot{\eta} \,,
 \label{susy12}\err
 where $\e$ is a real anticommuting parameter ($\e$ = $\e^*$).
 One way to describe these multiplets is via the superfields,
 \brr \Phi = \phi+i\,\t\,\psi
      \hspace{1in}
      \Lambda = \eta+\t\,B \,,
 \err
 where $\t$ is a real anticommuting coordinate.
 If we introduce superspace operators
 \brr Q &=&
      i\,\der/\der\t+\t\,\der_{\tau}
      \nonumber\\[.1in]
      D &=& i\,\der/\der\t-\t\,\der_{\tau} \,,
 \err
 then the transformation rules (\ref{susy12}) follow from
 acting on $\Phi$ and $\Lambda$ with
 $\d_Q(\e)=-i\,\e\,Q$.  We can write invariant actions as
 \brr S_\Phi &=&
      \int dt\,d\t\,\bpl\,\ft12\,\Phi\,\der_{\tau}\,D\,\Phi\,\bpr
      \nonumber\\[.1in]
      S_\Lambda &=&
      \int dt\,d\t\,\bpl\,-\ft12\,i\,\Lambda\,D\,\Lambda\,\bpr
 \err
 In terms of components, i.e., after performing the $\t$ integrations,
 there are described by
 \brr S_\Phi &=& \int dt\,\bpl\,
      \ft12\,\dot{\phi}^2
      -\ft12\,i\,\psi\,\dot{\psi}\,\bpr
      \nonumber\\[.1in]
      S_\Psi &=& \int dt\,\bpl\,
      -\ft12\,i\,\eta\,\dot{\eta}
      +\ft12\,B^2\,\bpr \,.
 \err
 It is also possible to add a superpotential for
 $\Phi$ by adding $\int dt\,d\t\,W(\Phi)$ to $S_\Phi$.
 Other interactions are also possible.

 \section{Automorphic Duality Transformations}
 \label{admaps}

~~~~ A useful operation which maps between the two irreducible
 $N=1$ multiplets was described in \cite{GatesLP}.
 In terms of components, this is realized by making the
 following replacements
 \brr (\,\dot{\phi}\,,\,\psi\,) &\leftrightarrow&
      (\,B\,,\,\eta\,) \,.
 \label{ad1}\err
 In terms of superfields, these are equivalent to
 \brr \Lambda &\leftrightarrow& -D\,\Phi \,.
 \err
 Under this operation, the transformation rules for the scalar and
 spinor multiplets are interchanged and the actions
 $S_\Phi$ and $S_\Lambda$ are also interchanged.  In other
 words, a map (\ref{ad1}) suffices to replace a scalar multiplet
 with a spinor multiplet and vice-versa.  Since this generates
 an automorphism on the space of superalgebra representations,
 this is referred to as an automorphic duality, or an AD map
 for short.  The term duality is used here not in the sense
 that the multiplets, or theories constructed using these
 multiplets, are equivalent.  Instead, the term implies
 simply that these constructions are paired by this operation.

 The AD map connecting a scalar multiplet with a spinor multiplet
 is intrinsically non-local.  This is because (\ref{ad1}) implies
 $\phi(\tau)\to \int\,dt\,B(\tau)$.  However, this is realized in a local way
on the transformation rules (\ref{rules}) and on the actions
(\ref{actionsm}) because $\phi$ always appears differentiated,
i.e., because there is a shift symmetry $\phi\to \phi+c$, where
$c$ is a constant parameter.  (A superpotential would generally
spoil
 this property.) It is possible to generalize these actions to describe
 quantum mechanical sigma models. In these cases, the presence
 of a shift symmetry implies that the target space has an isometry.
Interestingly, such isometries are precisely the ingredient needed
to couple a background vector field to the theory so as to switch
on a supersymmetry central charge \cite{Harmonic, Dualcc}.
 Therefore, the ability to perform automorphic duality transformations
 is equivalent to the ability to include a central term in the
 superalgebra. As shown in \cite{Harmonic}, these charges imply
 interesting target space dualities similar to $T$-dualities in string
 theory.  This motivates a basic connection between
 automorphic duality and non-trivial target space dualities.

 The AD map (\ref{ad1}) describes a quantum mechanical version of Hodge duality.
 To see this, note that in field theories Hodge duality
 maps a $P$-form $\Omega_P$ into a $D-P-2$ form $\tilde{\Omega}_{D-P-2}$
 via $d\,\Omega_P\to *\,d\,\tilde{\Omega}_{D-P-2}$. If one
 starts with a scalar field $\phi$ in one-dimension, then $D=1$ and
 $P=0$, in which case this implements a map
 $\phi\to\Omega_{-1}$, where $\Omega_{-1}$ is a formal
 ``minus-one"-form, an object
 whose exterior derivative is a zero-form.  This is precisely what
 characterizes the field $B$ which appears as the image of $\phi$ under
 (\ref{ad1}).

 \setcounter{equation}{0}
 \section{Extended Supersymmetry}
 \label{matter}

~~~~ The $N$-extended $d=1$ superalgebra is described by
 \brr [\,\d_Q(\e_1^I)\,,\,\d_Q(\e_2^J)\,]=
      -2\,i\,\e_1^I\,\e_2^I\,\der_{\tau} \,,
 \label{nextend}\err
 where $I=1,...,N$, and $\e_i^I$ are a set of real anticommuting
 parameters.  Although it is possible to include a central term on the
 right-hand side of (\ref{nextend}), we do not do so at this time
 \footnote{See Appendix A for a brief discussion pertaining to
 such central extensions.}. In this section we review a particular class of
 minimal representations to (\ref{nextend}).  These
 generalize the $N=1$ scalar and spinor multiplets described above.
 Many other representations exist which lie outside this
 class, however. Likely, all other representations can be
 discerned and organized using technology developed in
 \cite{GatesLP}.  In this and in the following three sections
 we briefly review these results, since these provide the
 mathematical basis behind the core presentation
 of this paper.

 \subsection{Scalar Multiplets}
~~~~ One class of representations describes generalized scalar
 multiplets.  The transformation
 rules are determined by making the following ansatz,
 \brr \d_Q\,\phi_i &=&
      -i\,\e^I\,(\,L_I\,)_i\,^{\hat{\jmath}}\,
      \psi_{\hat{\jmath}}
      \nonumber\\[.1in]
      \d_Q\,\psi_{\hat{\imath}} &=& \e^I\,
      (\,R_I\,)_{\hat{\imath}}\,^j\,\dot{\phi}_j \,,
 \label{rules}\err
 where $\phi_i(\tau)$ is a set of real commuting fields and
 $\psi_{\hat{\imath}}(\tau)$ is a set of real anticommuting fields.
 Ordinarily, supersymmetry requires an equal number of bosons and
 fermions.  Accordingly, the indices $i$ and $\hat{\imath}$ each have the
 same multiplicity, denoted $d$. Accordingly, $i=1,...,d$ and
 $\hat{\imath}=1,...,d$.  Furthermore, since $\phi_i$ and
 $\psi_{\hat{\imath}}$ are each real, it follows that the matrices
 $(\,L_I\,)_i\,^{\hat{\jmath}}$ and $(\,R_I\,)_{\hat{\imath}}\,^j$
 are real.
 The algebra (\ref{nextend}) imposes the following restrictions on
 $L_I$ and $R_I$,
 \brr (\,L_J\,R_I+L_I\,R_J\,)_i\,^j
      &=& -2\,\d_{IJ}\,\d_i\,^j
      \nonumber\\[.1in]
      (\,R_J\,L_I+R_I\,L_J\,)_{\hat{\imath}}\,^{\hat{\jmath}}
      &=& -2\,\d_{IJ}\,\d_{\hat{\imath}}\,^{\hat{\jmath}} \,.
 \label{lr}\err
 There is no reason from a purely algebraic point of view to
 impose an {\it a priori} relationship between $L_I$ and $R_I$.
 Nevertheless, a certain minimalist dynamical consideration
does imply one more restriction.  In particular, we require
 that the kinetic action described by
 \brr S_{SM}=\int dt\,\bpl\,
      \ft12\,\dot{\phi}^i\,\dot{\phi}_i
      -\ft12\,i\,\psi^{\hat{\imath}}\,\dot{\psi}_{\hat{\imath}}\,\bpr
      \,,
 \label{actionsm}\err
 be invariant under the transformations (\ref{rules}).
 In (\ref{actionsm}), indices are raised according to
 $\phi^i=\d^{ij}\,\phi_j$ and
 $\psi^{\hat{\imath}}=\d^{\hat{\imath}\hat{\jmath}}\,\psi_{\hat{\jmath}}$.
 The more general case, describing sigma models
 with a curved target space, involves additional subtlety not
 addressed in this paper.
 The action (\ref{actionsm}) is invariant under (\ref{rules}) only
 if
 \brr (\,L_I^T\,)^{\hat{\imath}j}=
      -(\,R_I\,)^{\hat{\imath}j} \,.
 \label{boo}\err
 This final requirement defines the operator
 $R_I$ in terms of $L_I$, or vice versa.  Taken together, the
 three requirements given in (\ref{lr}) and (\ref{boo}) describe
 an algebra which has been designated ${\cal GR}(d,N)$.

 It is possible to define ``twisted scalar multiplets"
 using the alternate transformation rules
 obtained by interchanging the placement of $L_I$ and $R_I$ in
 the transformation rules (\ref{rules}).  In this case, the
 algebraic requirements on $L_I$ and $R_I$ are identical to those
 in the untwisted case.

 \subsection{Spinor Multiplets}
~~~~ Another class of representations describes generalized spinor
 multiplets.  These are determined by analogy to the previous
 discussion. Accordingly, the transformation rules are determined using the
 following ansatz
 \brr \d_Q\,\eta_{\hat{\imath}} &=&
      \e^I\,(\,R_I\,)_{\hat{\imath}}\,^j\,B_j
      \nonumber\\[.1in]
      \d_Q\,B_i &=&
      -i\,\e^I\,(\,L_I\,)_i\,^{\hat{\jmath}}\,\dot{\eta}_{\hat{\jmath}}
      \,,
 \label{rulesfm}\err
 where $\eta_{\hat{\imath}}(\tau)$ describes $d$ real anticommuting
 fields and $F_i(\tau)$ describes $d$ real commuting fields.
 We require that the transformation rules (\ref{rulesfm})
 describe the algebra (\ref{nextend}).  We also impose
 that the minimalist kinetic action, given by
 \brr S_{FM}=\int dt\,\bpl\,
      -\ft12\,i\,\eta^{\hat{\imath}}\,\dot{\eta}_{\hat{\imath}}
      +\ft12\,B^i\,B_i
      \,\bpr \,,
 \label{actionfm}\err
 be a supersymmetry invariant.  Together, these imply
 precisely the same restrictions on $L_I$ and $R_I$ as given in
 (\ref{lr}) and (\ref{boo}).

 It is possible to define ``twisted spinor multiplets"
 using the alternate transformation rules
 obtained by interchanging the placement of $L_I$ and $R_I$ in
 (\ref{rulesfm}).  In this case, the
 algebraic requirements on $L_I$ and $R_I$ are once again identical to those
 in the untwisted case.

 \setcounter{equation}{0}
 \section{An Algebraic Basis for Generalized Superfields}

~~~~ The existence of the supermuliplets described above hinges on
 the ${\cal GR}(d,N)$ algebras, defined by
 \brr (\,L_I\,R_J+L_J\,R_I\,)_i\,^j
      &=& -2\,\d_{IJ}\,\d_i\,^j
      \nonumber\\[.1in]
      (\,R_I\,L_J+R_J\,L_I\,)_{\hat{\imath}}\,^{\hat{\jmath}}
      &=& -2\,\d_{IJ}\,\d_{\hat{\imath}}\,^{\hat{\jmath}}
      \nonumber\\[.1in]
      \d^{\hat{\imath}\hat{k}}\,(\,R_I\,)_{\hat{k}}\,^j &=&
      -\d^{jk}\,(\,L_I\,)_k\,^{\hat{\imath}} \,,
 \label{keys}\err
 where $(\,L_I\,)_i\,^{\hat{\jmath}}$
 and $(\,R_I\,)_{\hat{\imath}}\,^j$ describe two sets of $N$
 $d\times d$ matrices.  Let hatted indices take
 values in one vector space ${\cal V}_R\cong{\R}^d$ and let un-hatted indices
 take values in another vector space ${\cal V}_L\cong{\R}^d$.
 In this way, the $L_I$ matrices describe linear
 operators which map elements of ${\cal V}_R$ into
 elements of ${\cal V}_L$, and the $R_I$ matrices
 describe linear operators which map elements of ${\cal V}_L$ into
 elements of ${\cal V}_R$.
 It is useful to define four distinct sets of linear transformations
 which act on and between the
 two vector spaces ${\cal V}_L$ and ${\cal V}_R$ according to
 \brr  &&\{\,{\cal M}_L\,\}\,:\,
      {\cal V}_R\to{\cal V}_L
      \hspace{.3in}
      \{\,{\cal U}_L\,\}\,:\,
      {\cal V}_L\to{\cal V}_L
      \nonumber\\[.1in]
      &&\{\,{\cal M}_R\,\}\,:\,
      {\cal V}_L\to{\cal V}_R
      \hspace{.3in}
       \{\,{\cal U}_R\,\}\,:\,
      {\cal V}_R\to{\cal V}_R \,,
 \label{spaces}\err
 In this way
 $(\,L_I\,)_i\,^{\hat{\jmath}}\in \{\,{\cal M}_L\,\}$ and
 $(\,R_I\,)_{\hat{\imath}}\,^j\in \{\,{\cal M}_R\,\}$.
 Furthermore, $(\,L_I\,R_J\,)_i\,^j\in\{\,{\cal U}_L\,\}$ and
 $(\,R_I\,L_J\,)_{\hat{\imath}}\,^{\hat{\jmath}}\in
 \{\,{\cal U}_R\,\}$.  Each of the sets described in
 (\ref{spaces}) define a vector space in its own right.

 For a given value of $N$, there is a minimal value of $d$, called
 $d_N$, for which $N$ linearly independent real matrices $L_I$ exist which
 satisfy (\ref{keys}).  The value $d_N$ gives the
 number of off-shell bosonic (or fermionic) degrees of freedom in the minimal
 supersymmetry matter multiplets for that value of $N$.
 To determine $d_N$, notice that there
 is a unique way to write $N$ in terms of a mod 8 decomposition,
 $N=8\,m+n$.  Here $m=0,1,2,3,...$ counts cycles of 8, and $n=1,2,3,...$
 counts the position in the cycle.  For instance $N=7$ corresponds
 to $(m,n)=(0,7)$, $N=17$ corresponds to $(m,n)=(2,1)$, and
 $N=714$ corresponds to $(m,n)=(89,2)$. The values of $d_N$ are given by
 \brr d_N=16^m\,f_{\rm RH}(n) \,,
 \err
 where $f_{\rm RH}(n)$ is the so-called Radon-Hurwitz function
 \cite{PashTopp, GatesLP},
 defined as $f_{\rm RH}(n)=2^r$ where $r$ is the nearest integer
 greater than or equal to $\log_2{n}$.  The results are tabulated
 in Table \ref{dnvalues}.
  \begin{table}
 \begin{center}
 \begin{tabular}{|c||cccccccc|}
 \hline
 &\multicolumn{8}{|c|}{n}\\
 \hline
 m&1&2&3&4&5&6&7&8\\
 \hline
 \hline
 0&1&2&4&4&8&8&8&8\\
 1& 16&32&64&64&128&128&128&128\\
 2& 256&512&1024&1024&2048&2048&2048&2048\\
 3& 4096&8192&16,384&16,384&32,768&32,768&32,768&32,768\\
 & \multicolumn{8}{c|}{(etcetera)} \\
 \hline
 \hline
 type &
 N & AC & Q & Q & Q & AC & N & N \\
 \hline
 \end{tabular}\\[.1in]
 \caption{Values of $d_N$ where $N=8\,m+n$ for all
 $N\le 32$.  The $m=0$ row enumerates $d_1$ through $d_8$,
 the $m=1$ row enumerates $d_9$ through $d_{16}$, the
 $m=2$ row enumerates $d_{17}$ through $d_{24}$, and the
 $m=3$ row enumerates $d_{25}$ through $d_{32}$. This table
 can be continued to include an arbitrary number of rows.
 The final row indicates the ``type" of the ${\cal EGR}(d_N,N)$
 representations.}
 \label{dnvalues}
 \end{center}
 \end{table}
 Explicit matrix representations of $L_I$ and $R_I$ are given for
 $N\le 8$ in Appendix A of \cite{GatesRana1}.  This is
 generalized to arbitrary $N$ using a recursive scheme in
 \cite{GatesRana2}.

 The enveloping algebra
 ${\cal EGR}(d,N)\cong
 \{\,{\cal M}_L\,\}\oplus
 \{\,{\cal M}_R\,\}\oplus
 \{\,{\cal U}_L\,\}\oplus
 \{\,{\cal U}_R\,\}$
 consists of all linear maps on and
 between ${\cal V}_R$ and ${\cal V}_L$.
 Note that ${\cal GR}(d,N)\subset {\cal EGR}(d,N)$.
 A subalgebra of ${\cal EGR}(d,N)$ is generated by the
 two sets of $p$-forms defined as wedge products
 involving $L_I$ and $R_I$,
 \brr f_I &=& L_I
      \hspace{1.1in}
      \tilde{f}_{I}=
      R_I
      \nonumber\\[.1in]
      f_{IJ} &=& L_{[I}\,R_{J]}
      \hspace{.7in}
      \tilde{f}_{IJ}=
      R_{[I}\,L_{J]}
      \nonumber\\[.1in]
      f_{IJK} &=&
      L_{[I}\,R_{J}\,L_{K]}
      \hspace{.3in}
      \tilde{f}_{IJK}=
      R_{[I}\,L_J\,R_{K]}
 \err
 and so forth.  Each set of $p$-forms divides into
 even forms and odd forms, such that
 \brr f_{[{\rm odd}]}\oplus
      f_{[{\rm even}]}\oplus
      \tilde{f}_{[{\rm odd}]}\oplus
      \tilde{f}_{[{\rm even}]} \in
      \{\,{\cal M}_L\,\}\oplus
      \{\,{\cal U}_L\,\}\oplus
      \{\,{\cal M}_R\,\}\oplus
      \{\,{\cal U}_R\,\}
 \err
 Collectively, these operators generate
 an algebra denoted ${\bf \wedge} {\cal GR}(d,N)$.

 It is generally so that
 $\wedge {\cal GR}(d_N,N)\subset {\cal EGR}(d_N,N)$, although for
 some values of $N$, it turns out that
 $\wedge {\cal GR}(d_N,N)\cong {\cal EGR}(d_N,N)$.  In the latter case
 the algebra ${\cal EGR}(d_N,N)$ is said to be normal.  Otherwise
 the algebra falls into one of two classes, known as almost
 complex or quaternionic, depending on whether ${\cal EGR}(d_N,N)$
 contains two or four copies of ${\cal GR}(d_N,N)$, respectively.
 In the almost complex case, ${\cal EGR}(d_N,N)$ includes an
 operator, called ${\cal D}$ which interconnects the two copies of
 $\wedge {\cal GR}(d_N,N)$.  In the case of quaternionic algebras
 there is a triplet of operators ${\cal E}^{1,2,3}$ which
 interconnect the four copies of $\wedge {\cal GR}(d,N)$.
 In the balance of this paper we concern ourselves with
 constructions built using the algebras $\wedge {\cal GR}(d_N,N)$,
 rather than ${\cal EGR}(d_N,N)$.  A consequence is that
 the operators ${\cal D}$ and ${\cal E}^\a$ will not play a role
 in this paper.
 We suspect, however, that the operators ${\cal D}$ and ${\cal E}^\a$ will
 contribute in an interesting way in a more comprehensive supersymmetry
 representation theory. At the present time, however, their significance is
 not yet fully appreciated \footnote{One possibility is that these operators
 are needed to describe multiplets with a central charge.}.   We distinguish
 the vector spaces spanned by $\wedge{\cal GR}(d,N)$ by use of a
 ``prime" symbol.  For instance, $f_{[{\rm odd}]}\in \{\,{\cal M}_L\,\}{\bf '}$.
 The vector space $\{\,{\cal M}_L\,\}{\bf '}$ may be smaller than $\{\,{\cal
 M}_L\,\}$ in the case of almost complex or quaternionic algebras. Similar
 statements pertain to the other three vector spaces defined in (\ref{spaces}).

 In \cite{GatesLP, GatesRana1, GatesRana2} a close connection
 between the algebras ${\cal GR}(d_N,N)$ and $C(N,1)$ was
 exploited to describe the representation theory of the former
 algebra in terms of the representation theory of the latter.  This is
 helpful because Clifford algebra representations have been studied
 extensively, and are readily available in the literature. The same Clifford
 algebras play a seemingly different role in describing spinors in higher dimensional field theories.
 This may imply interesting ``shadow" relationships between representations of $D\ge 2$ superalgebras
 with analogous representations of $d=1$ superalgebras.

 A very brief synopsis of the connection between
 representations of ${\cal GR}(d,N)$ and those of $C(N,1)$ follows.
 For a more detailed description, the reader is referred to
 \cite{GatesLP} and references therein.  The Clifford algebra
 $C(N,1)$ is defined by
 \brr \{\,\Gamma_{\hat{I}}\,,\,\Gamma_{\hat{J}}\,\}=-2\,\eta_{\hat{I}\hat{J}} \,,
 \label{cn1}\err
 where $\hat{I}, \hat{J}=1,...,N+1$ and $\eta_{\hat{I}\hat{J}}={\rm diag}(1,...,1,-1)$.
 For each positive integer $N$ there exists a $2\,d\times 2\,d$
 matrix representation to (\ref{cn1}) such that
 the first $N$ Gamma matrices $\Gamma_I=\{\,\Gamma_1\,,...,\,\Gamma_N\,\}$
 are real and antisymmetric, and where
 \brr \Gamma_I=\ba{cc}0& L_I\\R_I & 0\ea \,.
 \err
 The smaller matrices $L_I$ and $R_I$ which appear here
 are each $d\times d$, and provide a representation of ${\cal GR}(d,N)$.

 \setcounter{equation}{0}
 \section{Clifford Algebra Superfields}

~~~~ The multiplets reviewed in section \ref{matter} arise from a
 derivation on a superspace ${\cal SM}\cong {\cal V}_L\oplus{\cal V}_R$,
 where ${\cal V}_L$ and ${\cal V}_R$ are the vector spaces described above.
 For instance, in the case of the scalar multiplet, $\phi_i(\tau)\in {\cal V}_L$ and
 $\psi_{\hat{\imath}}(\tau)\in {\cal V}_R$ are the superfield ``components".
 In this way, the world-line of a superparticle is described by a pair of trajectories,
 one in ${\cal V}_L$ and the other in ${\cal V}_R$.
 There are other possibilities, however.

 Consider instead a different superspace defined as
 ${\cal SM}'\cong {\cal U}_L\oplus{\cal M}_R$.
 Parameterize ${\cal SM}'$ using as component fields
 \brr \Phi_i\,^j(\tau) &\in& \{\,{\cal U}_L\,\}'
      \nonumber\\[.1in]
      \Psi_{\hat{\imath}}\,^j(\tau) &\in& \{\,{\cal M}_R\,\}' \,.
 \err
 Therefore, $\Phi_i\,^j(\tau)$
 and $\Psi_{\hat{\imath}}\,^j(\tau)$ describe fields on the particle world-line
 which take values in these vector spaces.
 We can expand the fields in terms of the bases
 $f_{[{\rm even}]}$ and $\hat{f}_{[{\rm odd}]}$ as follows,
 \brr \Phi_i\,^j &=&
      \phi\,\d_i\,^j
      +\phi^{IJ}\,(\,f_{IJ}\,)_i\,^j
      +\phi^{IJKL}\,(\,f_{IJKL}\,)_i\,^j
      +\cdots
      \nonumber\\[.1in]
      \Psi_{\hat{\imath}}\,^j &=&
      \psi^I\,(\,\hat{f}_I\,)_{\hat{\imath}}\,^j
      +\psi^{IJK}\,(\,\hat{f}_{IJK}\,)_{\hat{\imath}}\,^j
      +\cdots \,.
 \label{cliffsf}\err
 The pair $\{\,\Phi_i\,^j(\tau)\,,\,\Psi_{\hat{\imath}}\,^j(\tau)\,\}$ describe a Clifford algebraic superfield.
 The expansions (\ref{cliffsf}) terminate for any given finite value of $N$
 since any antisymmetric product with more than $N$ terms vanishes.
 (i.e., an $N$-form in $N$ dimensions is a top-form.)
 Define a supersymmetry transformation according to
 \brr \d_Q(\e)\,\Phi_i\,^j &=&
      -i\,\e^I\,(\,L_I\,)_i\,^{\hat{k}}\,\Psi_{\hat{k}}\,^j
      \nonumber\\[.1in]
      \d_Q(\e)\,\Psi_{\hat{i}}\,^j &=&
      \e^I\,(\,R_I\,)_{\hat{\imath}}\,^k\,\der_{\tau}\,{\Phi}_k\,^j \,.
 \label{tra}\err
 These rules automatically satisfy (\ref{basic12}) since by construction
 $L_I$ and $R_I$ obey (\ref{keys}).

 One can apply (\ref{tra}) to extract the transformation
 rules level-by level in the expansion (\ref{cliffsf}).
 This requires a careful use of the expressions in
 (\ref{keys}).
 In the general case, superfield transformation rules
 (\ref{tra}) imply the following rules for the level-expansion
 \brr \d\,\phi^{[p_{\rm even}]} &=&
      -i\,\e^{[I_1}\,\psi^{I_2\cdots I_p]}
      +(p+1)\,i\,\e_J\,\psi^{I_1\cdots I_p\,J}
      \nonumber\\[.1in]
      \d\,\psi^{[p_{\rm odd}]} &=&
      -\e^{[I_1}\,\dot{\phi}^{I_2\cdots I_P]}
      +(p+1)\,\e_J\,\dot{\phi}^{I_1\cdots I_P\,J} \,.
 \label{generic}\err
 Notice that the first term in $\d\,\phi^{[p]}$
 is a $p$-form obtained as a wedge-product between the one-form parameter $\e^I$
 and a fermionic $(p-1)$-form.  In the case $p=0$ this term vanishes
 because $(p-1)<1$, and therefore there is no corresponding fermion.

 A word on terminology.  In traditional superfields ${\cal S}={\cal S}
 (\,t\,,\,\t^1\,,...,\,\t^N\,)$ one refers to the sequence of component fields in
 terms of ``lowest component" to ``highest component" where, roughly speaking,
 the component number corresponds to the associated power
 of $\theta^I$ which appears in a formal Taylor series expansion of
 ${\cal S}$.  In Clifford algebra superfields we refer to the
 analogous
 sequence using the terms ``level-zero" to ``level-N". In this case the
``level" corresponds to the $\wedge {\cal GR}(d_N,N)$ form-degree of
 the terms in question.  For each choice of $N$ there are two distinct
 Clifford algebra superfields.  One has a level-zero boson and one has
 a level-zero fermion.  We refer to the former as a bosonic Clifford algebra superfield
 and to the latter as a fermionic Clifford algebra superfield.
 The bosonic Clifford algebra superfield is also called the ``base superfield" for the
 corresponding value of $N$.  In the case of a bosonic Clifford
 algebra superfield, the even levels are described by the field $\Phi_i\,^j
 (\tau)$ and the odd levels are described using the field $\Psi_{\hat{
 \imath}}\,^j(\tau)$.  For instance, the $N=3$ base superfield has a level-zero boson
 $\phi$, three level-one fermions organized as a vector $\psi^I$,
 three level-two bosons organized as a two-form $\phi^{IJ}$ , and one level-three
 fermion organized as a three-form $\psi^{IJK}$.  The three form is equivalently
 described as a one-form in terms of $(\,*\psi\,)^I=\ve^{
 IJKL}\,\psi_{JKL}$.

 \setcounter{equation}{0}
 \section{Root Superfields}
 \label{dualq}

~~~~ Clifford algebraic superfields describe only a restricted class of
 multiplets.  Moreover, for the cases $N\ge 4$ these
 representations are reducible. This construction complements
 the superfields described previously
 using elements of ${\cal V}_L\oplus {\cal V}_R$ as component fields.
 Nevertheless, these two sorts of superfields do not yet provide a
 sufficient basis for a comprehensive representation theory.
 A big step in that direction is obtained by using the Clifford
 algebraic superfields as a ``base" upon which a variety of
 operations can be performed so as to obtain a much larger class
 of representations.

 Take a Clifford algebraic superfield (\ref{cliffsf}), and write
 the components as
 \footnote{The convention used here is slightly different than the
 convention defined in \cite{GatesLP}.  In this modified \newline
$~~~~~$ convention, the {\it odd} labels $a_{i={\rm odd}}$ in the base
 superfield differ by a minus sign as compared to \newline
$~~~~~$  that paper.}
 \brr (\,\phi\,,\,\psi^I\,,\,\phi^{IJ}\,,...\,)
      &=& (\,\der_{\tau}^{-a_0}\,\tilde{\phi}\,,\,
      \der_{\tau}^{a_1}\,\tilde{\psi}^I\,,\,
      \der_{\tau}^{-a_2}\,\tilde{\phi}^{IJ}\,...\,) \,,
 \label{beat}\err
 etcetera, where $a_i\in\Z$.  For the case where all of the $a_i$
 are zero, the components $(\tilde{\phi}\,,\,\tilde{\psi}^I\,,\, \tilde{
 \phi}^{IJ}\,,...\,)$ describe the base superfield.  However, when at
 least one of the labels is non-zero, then the structure of the superfield changes
 in an interesting way.  For instance, when one of the bosonic labels is $1$, this
 means that the corresponding component is written
 as the anti-derivative of a ``dual" component.  To be more concrete, if $a_2=1$ this
 would imply  that $\phi^{IJ}(\tau)=\int^\tau d\tilde{\tau}\,\tilde{
 \phi}^{IJ}(\tilde{\tau})$ or, equivalently, that $\der_{\tau}\,{\phi}^{IJ}
 =\tilde{\phi}^{IJ}$.  Note that this describes an automorphic
 duality transformation.  The relationship between the base
 mutliplet and a generic root multiplet is described in terms
 of sequences of AD maps.

 It is also important to realize that the usual level of a component
 field in the conventional superspace approach is no longer rigidly
 linked to the order of the Clifford algebra elements when at least
 one of the exponents is non-vanishing.

 The root superfields utilizing various choices of $a_i$,
 in general describe distinct representations of supersymmetry.
 It is useful to invent a nomenclature to refer to these.  Accordingly, we
 designate the base multiplet, where all of the $a_i$ vanish,
 using a so-called root label $(0...0)_+$, which includes $N+1$ zeros.
 The subscript $+$ designates that the zero form is a boson.
 In the case where the zero form is a fermion, the corresponding
 root-label is $(0...0)_-$.  Starting with the base superfield,
 another superfield is obtained by dualizing on one of the component
 levels. For instance, if we started with the $N=3$ base
 superfield $(0000)_+$ and dualized at level-two, i.e., dualized the
 two-form $\phi^{IJ}$, then we would obtain the superfield
 $(0010)_+$.  Other cases are labelled similarly.

 At each value of $N$ the base superfield $(0...0)_+$, plays a
 special role in the representation theory.  It proves helpful to
 give this multiplet the special symbol $\Omega_{0+}^{(N)}$.
 Similarly, we denote the Clifford algebraic superfield having
 a fermionic zero-form, i.e., $(0...0)_-$, as $\Omega_{0-}^{(N)}$.

 The numbers $a_i$ in the root superfield label $(a_0,...,a_{N})_\pm$
 can take on any integer value.  However, the multiplets for which
 $a_i\in\{\,0\,,\,1\,\}$ are of particular interest.  We refer to the set of such
 multiplets as the ``root tree".  In these cases, the label can be read as a
 binary number.  For instance, the sequence of numbers in the label
 $(0101)_+$ can be read as $0\cdot 2^3+1\cdot 2^2+0\cdot 2^1+1\cdot
  2^0=5$.  We therefore denote this multiplet using the notation
 $\Omega_{5+}^{(3)}$.  In this way, we can describe a useful class
 of multiplets using the concise names $\Omega_{\mu\pm}^{(N)}$,
 where $\mu$ are integers such that $0\le \mu\le(\,2^{N+1}-1\,)$.
 As it turns out, there is in general much redundancy in these names.
 For instance, for each choice of $N$, the multiplets in the root tree having
 root label $(\,1\,,\,a_1\,,...,\,a_N\,)_\pm$ are the same as the multiplets
 $(\,0\,,\,1-a_1,\,...,\,1-a_N\,)_\pm$.  Thus, without loss of generality we consider $0\le \mu\le(\,2^N-1\,)$.

 \setcounter{equation}{0}
 \section{Multiplet Adinkras}
 \label{diagrams}

~~~~ In this section we define a powerful diagrammatic technique
 which usefully encodes many aspects of supersymmetry multiplets.
 According to this scheme, each multiplet has a corresponding
 distinctive symbolic form, which we refer to as an
 adinkra symbol, or an adinkra for short. An adinkra uses white circles
 to represent bosons and black circles to represent fermions.
 In either case the circles are called nodes.
 The nodes are interconnected using oriented line segments, referred to as
 arrows.  The arrows are directed from nodes representing lower
 component fields toward nodes representing higher component
 fields.

 \subsection{$N=1$ Adinkras}
 \label{n1sec}
~~~~ Each of the two irreducible $N=1$ multiplets
 include off-shell one bosonic and one fermionic
 degree of freedom.  Accordingly, the adinkras for these multiplets
 include one white node, to represent the boson, and one black node,
 to represent the fermion.  In the case of the scalar multiplet,
 the boson is the lower component and the fermion is the higher
 component.   Accordingly, the adinkra for the scalar multiplet is
 \begin{center}
 \includegraphics[width=.25in]{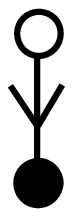} \,.
 \end{center}
 Since the arrow points toward the black node, it is clear
 that the fermion is the higher component in this multiplet.
 We can use the adinkra as a method for identifying this
 multiplet.  We recall, however, that the scalar multiplet can also be
 described with a root label, as $(00)_+$, or
 using Omega notation, as $\Omega_{0+}^{(1)}$. Each of these
 three schemes has advantages and disadvantages.
 In the balance of this paper we
 demonstrate how the adinkra is useful for organizing the
 assembly of $N=1$ multiplets into higher-$N$ multiplets,
 for identifying irreducible multiplets, and for describing
 gauge invariance. We will use root labels or Omega notation in cases where
 these choices are advantageous, however, since the three notational
 schemes usefully complement each other.

 In the case of the spinor multiplet the fermion is the lower
 component and the boson is the higher component.  Accordingly,
 the adinkra for the spinor multiplet is
 \begin{center}
 \includegraphics[width=.25in]{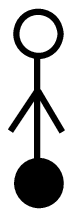} \,.
 \end{center}
 Since the arrow points toward the white node, it is clear that
 the boson is the higher component in this multiplet.
 The spinor multiplet can also be described with a root label, as $(00)_-$, or
 using Omega notation, as $\Omega_{0-}^{(1)}$.

 The adinkras symbolically encode the supersymmetry transformation rules for the
 corresponding multiplets.  This is seen easily in the $N=1$ case by comparing the
 adinkras shown above with the transformation rules given in
 (\ref{susy12}).  The supersymmetry transformation rule for a
 generic component field $f(\tau)$ corresponding to an adinkra node is given
 in this case by
 \brr \d_Q(\e)\,f=\pm i^b\,\e\,\der_{\tau}^\lambda\,f \,,
 \label{trule}\err
 where $b=1$ for bosons and $b=0$ for
 fermions, and $\lambda=1$ for lower components and $\lambda=0$
 for higher components. The ambiguous sign appearing in this
 rule must be chosen identically at each of the two nodes.
 The choice of which sign is irrelevant in the $N=1$ case, since
 this can be flipped by redefining either node with a multiplicative minus sign
 \footnote{We refer to nodes and to the corresponding
 component fields as if they were the same \newline
$~~~~~~$  entity.  Thus,
 by scaling a node we mean that we are scaling the corresponding
 field.}.
 We refer to this sign choice as the
 ``arrow parity". The concept of arrow parity becomes important when
 combining $N=1$ multiplets to form higher-$N$ multiplets,
 as we explain below. The reader is encouraged to derive the
 transformation rules (\ref{susy12}) from the two $N=1$ adinkras
 presented above.  This is a simple exercise which illustrates
 only a part of the hidden meaning in these symbols.  The useful
 mnemonic is that each boson receives a factor of $i$ in its
 transformation rule, and higher components appear differentiated.
 This rule generalizes to generate the transformation rule
 corresponding to any arrow in any of the adinkras in the root tree
 for any value of $N$, but requires modification for adinkras
 not in the root tree.

 An adinkra symbol does not have an intrinsic orientation; either of
 the adinkras shown above can be rotated arbitrarily in the plane of the page.
 For certain purposes, it is useful to draw the symbol in a
 particular manner, however.  For instance, it is conventional to present
 transformation rules starting with the lowest component
 at the top of a list, and work toward the highest component at the
 bottom of a list.
 The adinkra most faithfully represents this
 structure if all arrows point downward.  This was the choice made
 in the case of the scalar adinkra, shown above, but not in the
 case of the spinor adinkra.  The choice made in the case of the
 spinor adinkra serves a different purposes, as will become clear.
 Since arrows typically point from lower components to higher
 components we refer to a node as being ``higher" than an adjacent
 node if an arrow points from the former node toward the latter
 node.  For multiplets in the root tree all nodes conform to an
 unambiguous hierarchy such that each node is either higher,
 lower, or at the same height as each of the other nodes.
 There are interesting other adinkras, not in the root tree, for which there is
 not an unambiguous hierarchy.

 Recall that the scalar multiplet can be mapped into the spinor
 multiplet using an AD map.  The effect of this map is to exchange
 the roles of which of the two adjacent nodes is higher or lower.
 Accordingly, this map can be visualized as a reversal
 of the ``sense" on the arrow connecting the two adjacent nodes
 in the adinkra,
 \begin{center}
 \includegraphics[width=.9in]{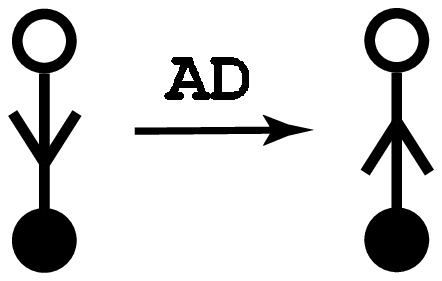} \,.
 \end{center}
 We can interpret the boson in the scalar multiplet as the
 lowest level in the Clifford algebra superfield
 $(00)_+$.  The boson in the spinor multiplet can be
 interpreted as the lowest level in the root superfield $(01)_+$.
 Thus, if the adinkra is oriented in such a way that the superfield
 levels are manifest, then by representing the AD map as an operation
 which reverses the sense of an arrow, but which leaves the position
 of the nodes unchanged, we faithfully preserve the manifestation
 of levels on the structure of the adinkra.

 When we implement the map described above, $(00)_+\to (01)_+$ we have
 replaced the level-one fermion in $(00)_+$ with a dual component.
 Accordingly, we say that we have ``dualized" at level-one.
 This is readily visualized not only on the adinkra, but
 also on the root label, since level-one corresponds to the
 second index in the root label.  The rule for implementing
 AD maps on adinkra symbols is that dualizing at a certain
 level corresponds to flipping all arrows with connect to nodes
 at that level.

 Suppose we implement a different AD map, this time by
 dualizing at level-zero.  This corresponds to $(00)_+\to (10)_+$,
 since level-zero corresponds to the first index in the root label.
 Using the rules described above, we represent this
 by reversing the sense of every arrow connecting to the top node
 in the $(00)_+$ adinkra.  This produces the
 same diagram obtained by our previous duality operation.
 In other words, the adinkra for $(10)_+$ is the same as
 the adinkra for $(01)_+$.  Thus, the $N=1$ spinor multiplet
 is described by the equivalent labels $(01)_+\cong (10)_+$.

 The $N=1$ spinor multiplet is also described by the label
 $(00)_-$, since it corresponds to the $N=1$ Clifford algebra superfield
 having a level-zero fermion.  Using this label, an AD map could
 be performed at level-one as $(00)_-\to (01)_-$.  This is
 implemented by reversing the sense of every arrow which connects
 to the level-one node in the spinor multiplet adinkra.
 Each of the AD maps described so far merely
 toggle between the two possible $N=1$ adinkras.  We discover in this way
 a nexus of congruencies in the root labels. Specifically, $(00)_+\cong (01)_-\cong
 (10)_-$ and $(00)_-\cong (01)_+\cong (10)_+$.  In terms of Omega
 notation, this result corresponds to
 $\Omega_{0+}^{(1)}\cong\Omega_{1-}^{(1)}\cong\Omega_{2-}^{(1)}$
 and
 $\Omega_{0-}^{(1)}\cong\Omega_{1+}^{(1)}\cong\Omega_{2+}^{(1)}$.
 In the $N=1$ case AD maps comprise
 ${\Z}_2$ generators which link the two
 congruency classes.

 There is another useful ${\Z}_2$ map which is distinct from the
 AD maps described so far.  This second map is described
 by replacing every bosonic node in a given adinkra with a fermionic node, and vice-versa.
 This operation was introduced in \cite{GatesKetov}, where it was
 deemed a Klein flip.
 In the case of $N=1$ supersymmetry a Klein flip
 has the same effect as an AD map, since it toggles
 between the two adinkras.  The circumstance that AD maps
 and Klein flips generate indistinguishable automorphisms
 is special to the case $N=1$ where it is a consequence of the
 relative simplicity of the space of irreducible representations.

 \setcounter{equation}{0}
 \section{$N=2$ Adinkras}

 ~~~~ Consider the $N=2$ scalar multiplet described by (\ref{rules}).
 To be specific, chose the ${\cal GR}(2,2)$ matrices
 according to $L_1=R_1=i\,\s_2$ and
 $L_2=-R_2=-{\mathbb I}_2$, where ${\mathbb I}_d$ is the $d\times d$ unit matrix.
 This describes the unique representation of ${\cal GR}(2,2)$.
 It is easy to translate the corresponding transformation rules
 into an adinkra symbol using the rules described above.  The
 result is
 \begin{center}
 \includegraphics[width=.7in]{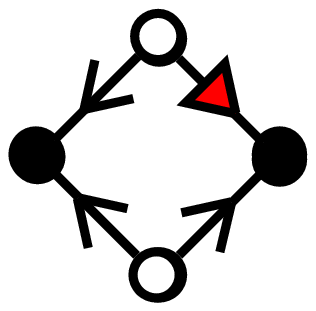} \,,
 \end{center}
 where we have distinguished one of the arrows for a reason to be
 explained shortly.  But first, we explain the general structure
 of this adinkra.  Each pair of parallel arrows corresponds to
 one of the two supersymmetry transformations.  For the sake of
 concreteness, lets say that the red arrow and the arrow opposite
 the red arrow correspond to the first supersymmetry, described by
 parameter $\e^1$, and that the remaining two arrows correspond to
 the second supersymmetry, parameterized by $\e^2$.

 The reader is
 encouraged to use (\ref{rules}) along with the representation of
 ${\cal GR}(2,2)$ given above, to verify the rule (\ref{trule})
 node-by-node and arrow by arrow.  To do this, let the top node
 represent $\phi_1$ and the bottom node represent $\phi_2$,
 and let the left node represent $\psi_{\hat{1}}$ and the right node
 represent $\psi_{\hat{2}}$. This exercise will expose the
 special characteristic which distinguishes the red arrow in this
 diagram.  Namely, this arrow corresponds to a choice
 of minus sign in (\ref{trule}), whereas the remaining three arrows
 correspond to a choice of plus sign in this rule.
 This indicates a topological characteristic
 required of any adinkra symbol for the cases $N\ge 2$, as explained
 presently.  Each of
 these symbols has closed circuits which can be traced on the
 diagram by following arrows from node to node.  A consequence of
 the minus signs in the first two equations in (\ref{keys}) is that the
 sum of the arrow parities associated with any four-node closed
 circuit must be negative
 \footnote{Note that there is no correlation between the parity of an arrow
 and its orientation.}.

 It is possible, of course, to redefine any component field
 by use of a multiplicative minus sign.
 We refer to this benign operation by saying that we have flipped the sign
 of a node.  Notice that by flipping a sign on either of the nodes
 adjacent to the negative-parity arrow, the position of the
 negative parity arrow shifts around the diagram. Another
 possibility is to flip the sign on one of the nodes not
 adjacent to the negative parity arrow.  The effect of this
 is to grow two more negative parity arrows.  Note that in the
 case of the square diagram flipping node signs necessarily
 changes the parity of exactly two arrows.  In this way the
 sum rule is preserved under such operations, although the
 parity of any given arrow can be flipped by
 field redefinitions.

 The reader might imagine that keeping proper track of arrow
 parities could become a complicated business in higher-$N$
 diagrams.  Fortunately, there is a simple algorithm which
 handily takes care of this for us in many, if not all,
 circumstances.  This algorithm relies on
 the root superfields described in section \ref{dualq},
 each of which is derived from the base superfields
 by AD maps and Klein flips.  The arrow parities for the
 base superfields are consistently
 dictated by the transformation rules given
 in (\ref{generic}).  As a result, one can draw the
 adinkra symbol for a base superfield without specifying
 the arrow parities, knowing that these can be chosen in
 a consistent manner.  It is then possible to derive a
 variety of related multiplets by implementing AD maps (by flipping
 arrows) and Klein flips (by flipping node colors), again without
 regard for arrow parity, since consistency is ensured by the
 fact that the base adinkra is consistent by construction.

 It is a noteworthy fact that the $N=2$ scalar multiplet described
 above is, in fact, the $N=2$ base multiplet $\Omega_{0+}^{(2)}$.
 This can be verified by determining the transformation rules for
 $\Omega_{0+}^{(2)}$ using (\ref{generic}), and then translating
 these into an adinkra symbol.  The reader is encouraged to do
 this.

 We are now in a position to describe the $N=2$ root tree
 using adinkra symbols.
 Consider the following four $N=2$ adinkras,
 \begin{center}
 \includegraphics[width=3in]{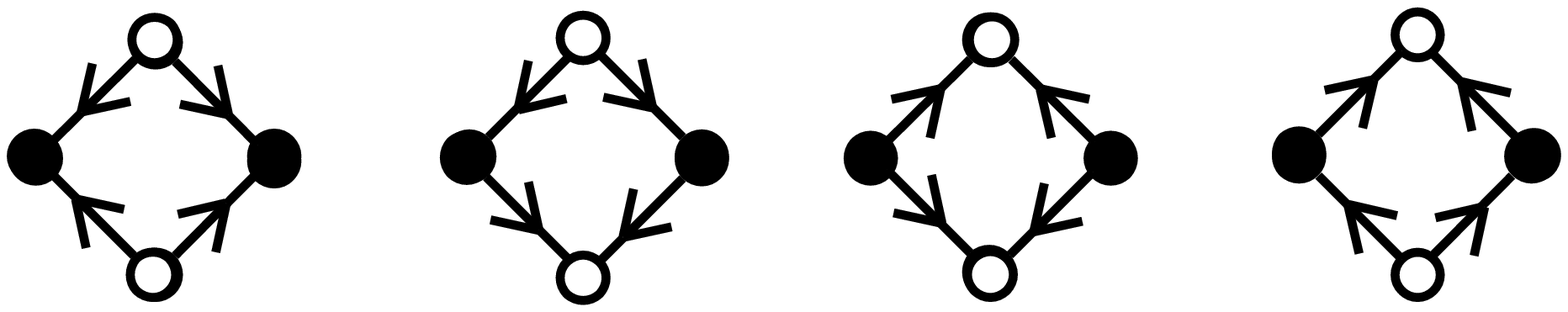} \,.
 \end{center}
 The first of these is the base adinkra $(000)_+$ which
 we have described at length already.  By convention, we have suppressed any special
 markers indicating arrow parity, since as explained above these
 are not necessary.  We have oriented this adinkra such that the top node
 corresponds to level-zero, the middle nodes correspond to level-one,
 and the bottom node corresponds to level-two. The second of the adinkras shown here
 is obtained from the first by dualizing at level-two.  This is
 clear because the relationship between the second adinkra and the
 first is that both arrows which connect with the level-two node
 have been flipped. This second multiplet has root label $(001)_+$
 and Omega designation $\Omega_{1+}^{(2)}$.  The third adinkra is
 obtained from the first by flipping all arrows which connect to
 level-one nodes.  (In this case this describes all arrows!)  Thus
 the third multiplet has root label $(010)_+$ and Omega designation
 $\Omega_{2+}^{(2)}$.  The fourth adinkra is obtained from the
 first by flipping both arrows which connect with the level-zero
 node.  Thus, this multiplet has root label $(100)_+$.  This
 fourth adinkra is also obtained by flipping all arrows connecting
 to the level-one nodes and then flipping both arrows connecting to
 the level-two nodes.  According to this second interpretation the
 fourth adinkra has the equivalent root label $(011)+$ and Omega designation
 $\Omega_{3+}^{(2)}$.  The three distinct adinkras
 $\Omega_{\mu+}^{(2)}$ where $\mu=0,1,2$, form a sequence,
 which we refer to as the ``base sequence" for $N=2$.

 The $N=2$ adinkras presented so far describe all of the
 adinkras which can be obtained from the base adinkra by
 AD maps. Notice that the $(001)_+$ adinkra is homologous to the
 $(100)_+\cong (011)_+$ adinkra.  This is seen by rotating
 either of these by 180 degrees. Thus, $(001)_+\cong(100)_+$,
 or equivalently $\Omega_{3+}^{(2)}\cong\Omega_{1}^{(2)}$.
 Thus, the number of multiplets in the AD orbit connected to the
 $N=2$ base multiplets is three, not four.

 There is still another multiplet in the $N=2$ root tree
 yet to be described.
 To locate this missing multiplet, consider that
 multiplet obtained from the base mutliplet by implementing a Klein flip,
 namely $(000)_-$.  Consider as well the set of multiplets
 connected with this one via AD maps.  This set is
 described by the following four adinkra symbols,
 \begin{center}
 \includegraphics[width=3in]{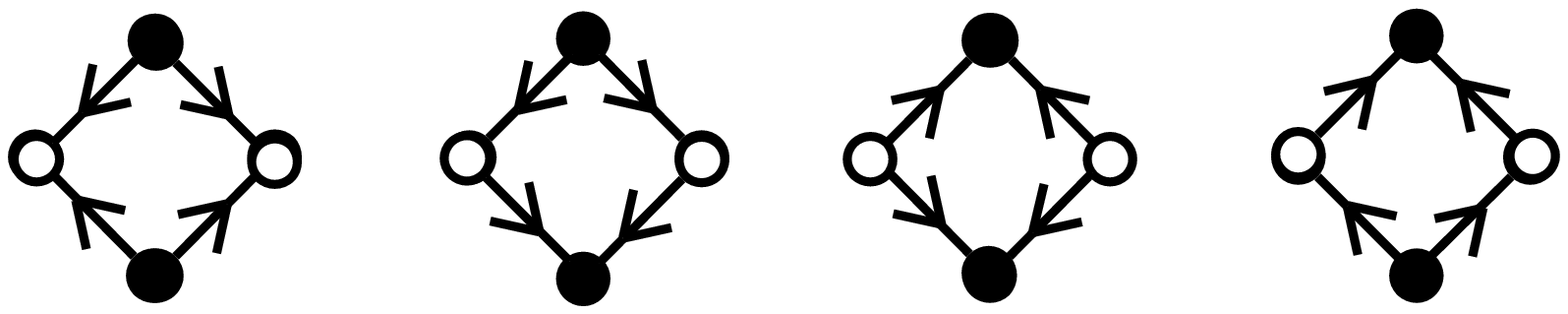}
 \end{center}
 The first of these is the image of the base adinkra under
 a Klein flip, i.e., $(000)_-$.  The other three adinkras shown here are
 obtained from this one by dualizing at levels two, one
 and zero, respectively.  Accordingly, these
 describe the respective multiplets $(001)_-$, $(010)_-$ and
 $(100)_-\cong (011)_-$ or, using Omega notation,
 $\Omega_{1-}^{(2)}$, $\Omega_{2-}{(2)}$ and $\Omega_{3-}^{(2)}$,
 again respectively.  This sequence of four adinkras
 describes the respective images under Klein flips of the base sequence
 shown previously.  This is easily verified by
 flipping the color of all nodes and then comparing these
 two sequences.
 By rotating the $\Omega_{3-}^{(2)}$ adinkra by 180 degrees, we
 observe that this is homologous to the $\Omega_{1-}^{(2)}$
 adinkra.
 The three distinct adinkras
 $\Omega_{\mu-}^{(2)}$ where $\mu=0,1,2$, form a sequence,
 which we refer to as the ``mirror sequence" for $N=2$.

 We now see more congruencies, this time showing equivalences
 between elements of the base sequence with elements in the mirror
 sequence.  For instance, by rotating the $\Omega_{0-}^{(2)}$
 adinkra by 90 degrees, we see that this is identical to the
 $\Omega_{2+}^{(2)}$ adinkra.  Also, by rotating the
 $\Omega_{2-}^{(2)}$ adinkra by 90 degrees we see that this is
 identical to the $\Omega_{0+}^{(2)}$ adinkra.  Therefore, the
 only adinkra which appears in the mirror sequence which is
 distinct from all of those in the base sequence is
 $\Omega_{1-}^{(2)}$.  In this way we determine that the $N=2$ root
 tree has four elements, which can be described as $\Omega_{0\pm}^{(2)}$ and
 $\Omega_{1\pm}^{(2)}$.

 \subsection{Adinkra Folding}
~~~~ We have seen that the irreducible $N=1$ adinkra symbols each
 comprise two nodes connected by an oriented line segment.  Thus, these
 symbols span only one linear dimension.  By way of contrast, the
 irreducible $N=2$ adinkra symbols comprise four nodes configured
 at the corners of a square, which has oriented line segments as
 edges.  Thus, the $N=2$ adinkras span two dimensions.  The reason
 for this is that the two supersymmetries are represented by
 orthogonal arrows.  Following this logic, the $N=3$ adinkras
 span three dimensions, so that the three supersymmetries can be
 represented using three mutually-orthogonal sets of arrows.
 As $N$ increases this leads to complicated symbols which would
 be difficult to render on a page.  However, there is a useful
 operation one can perform on adinkras, which allows the drawing of
 many of these for any $N$ as a linear chain, yet retains the
 full symbolic power.  In this subsection we describe
 this process for the case $N=2$, although the technique
 generalizes to higher $N$.

 We can squash or fold any adinkra by moving bosonic nodes onto other
 bosonic nodes and at the same time moving fermionic nodes onto
 other fermionic nodes, while simultaneously maintaining all
 arrow-node connections. This can be done provided that all arrows land on
 identically oriented arrows.
 For instance, consider the $\Omega_{1+}^{(2)}$ adinkra
 shown above. In this case, we can pinch the two fermionic nodes
 together, as follows,
 \begin{center}
 \includegraphics[width=2.5in]{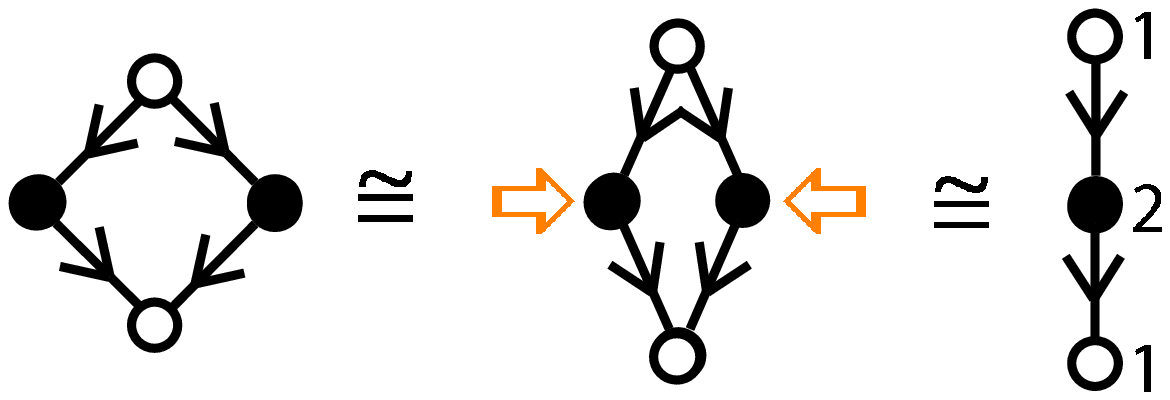} \,.
 \end{center}
 At the end of the process, the two fermionic nodes are
 coincident.  We indicate the multiplicity of this
 compound node by placing a numeral $2$ next to the node.
 Nodes representing  more degrees of freedom occur
 in more complicated adinkra symbols.
 The node multiplicity is indicated by a numeral
 placed adjacent to the node.

 In many cases, the arrows in an adinkra symbol are structured in such a way
 that permits additional folds after the first one.  Consider, for
 instance the $N=2$ base adinkra, which can be folded as follows,
 \begin{center}
 \includegraphics[width=4in]{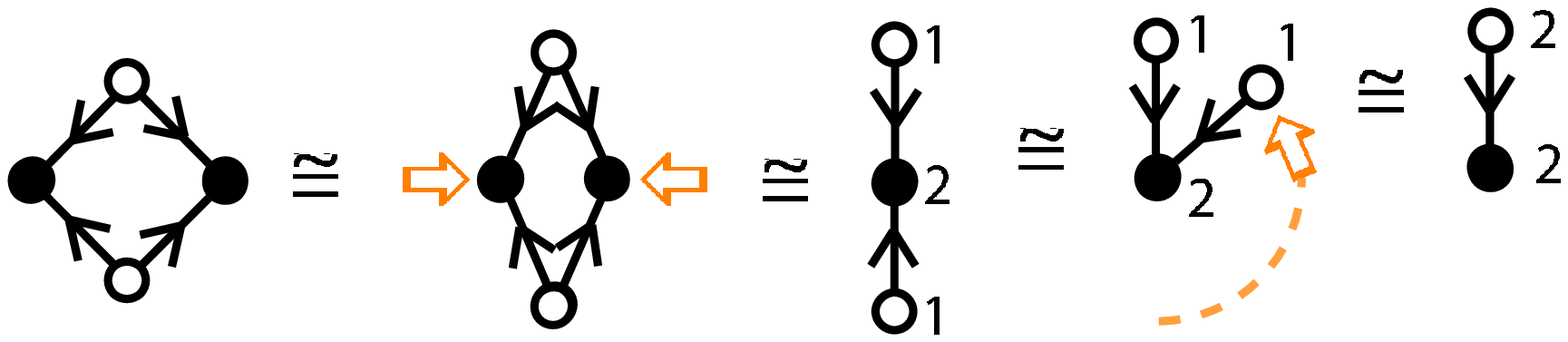} \,.
 \end{center}
 In this case we begin by pinching the two fermionic nodes
 together in a manner identical to the operation performed
 above on the $\Omega_{1+}^{(2)}$ adinkra.  In this case,
 since all the arrows continue to point to the multiplicity-two fermionic
 node, it is possible to swivel the bottom bosonic node, using
 the compound fermionic node as a pivot,
 until it coincides with the top bosonic node.  In this way we obtain
 a folded form which involves two multiplicity-two nodes,
 one bosonic and one fermionic, connected by one arrow which now
 represents both supersymmetries.

 By using similar folding operations, all of the elements of the
 root tree for any value of $N$ can be arranged into a linear chain.
 Many other adinkras, which are not elements of the root tree
 can not be folded into a linear chain;
 these describe an interesting class of multiplets which is described
 in the following section.
 Each distinct adinkra has a fully-folded form which is
 distinct from the fully-folded forms of all other distinct
 adinkras.
 It is possible to identify each distinct
 supersymmetric multiplet with a unique fully-folded adinkra
 symbol.  The folded adinkras can be unfolded, using certain
 rules, in such a way as to recover the fully unfolded adinkra.
 As we explain below, it is often useful to start with a
 fully-folded adinkra symbol, and then only partially unfold
 this before implementing duality maps, by making arrow reversals.
 This can then be re-folded to obtain a new adinkra.

 \setcounter{equation}{0}
 \section{Escheric Multiplets}
 \label{esmults}

 ~~~~In this section we describe some
 surprising unanticipated aspects of supersymmetry
 representations which become evident when this subject is
 structured in terms of adinkra symbols.  We present these
 observations at this point, immediately following our description
 of the basic $N=2$ adinkras, because these aspects are most clearly
 illustrated in the context of $N=2$ supersymmetry. We continue the main thrust
 of the paper, by generalizing our technology to the cases $N=3$
 and $N=4$, in the sections which follow this one.

 By including AD maps and Klein flips together,
 we are able to effectively realize dualities
 on the $N=2$ base adinkras node-by-node rather than
 level-by-level. To illustrate this, start with the
 $N=2$ base adinkra, perform a Klein flip, then dualize
 at level-two in the resulting adinkra, then rotate the
 adinkra by 90 degrees,
 \begin{center}
 \includegraphics[width=4in]{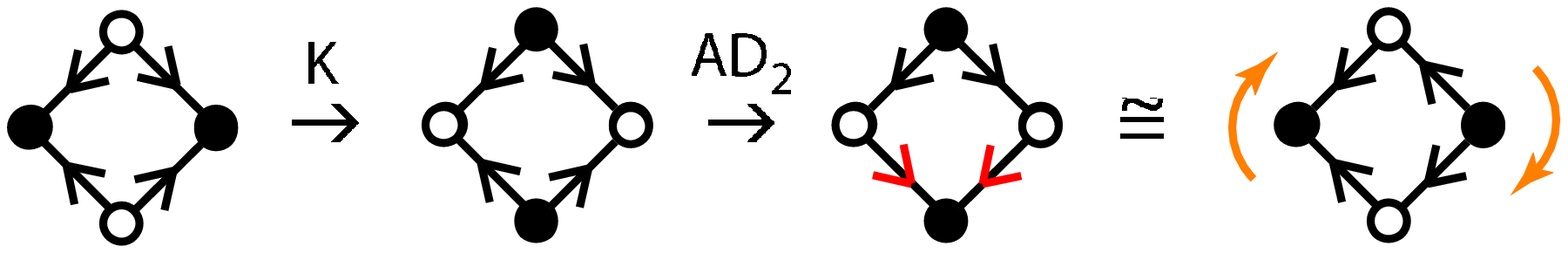} \,.
 \end{center}
 The result of this sequence of operations is the same as if
 we dualized on only one of the two level-one fermion nodes
 in the base adrinkra.  In other words, this operation is
 equivalent to
 \begin{center}
 \includegraphics[width=2in]{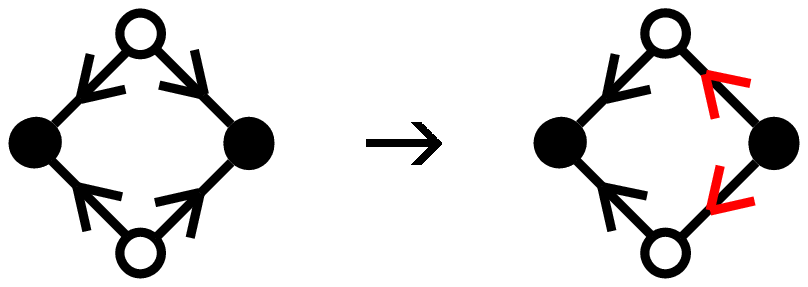} \,.
 \end{center}
 This begs the question as to whether we can realize
 dualities node-by-node rather than level-by-level
 as a general rule.  The answer is that, generally,
 such transformations cannot be implemented by a combination
 of Klein flips and AD maps.  The reason for this is connected
 to the fact that the AD maps, as described
 above, act in a strictly level-specific manner on the
 root superfields.
 The case of the $N=2$ base adinkra provides an exception,
 as we have seen.  Is it nevertheless possible to implement dualities
 node-by-node on any given adinkra?  Does this supply us with
 new representations of supersymmetry?  The answer
 to both questions is, interestingly, yes.
 These operations generally produce new multiplets
 which lie outside of the root tree and which have
 noteworthy nontrivial topological features.  It is also
 possible in this way to obtain multiplets which represent
 centrally-extended superalgebras.

 The simplest example of this phenomenon is described by starting
 with the $\Omega_{2+}^{(2)}$ adinkra and dualizing at one of the
 two level-one fermionic nodes.  As it turns out, there are two
 rather different senses to interpret ``dualization" in this
 context. The first sense is to simply reverse the sense of each
 arrow which connects to the node being dualized.
 As we will see, in the current context
 this process produces a rather different sort
 of multiplet, one which does not strictly represent the superalgebra
 (\ref{nextend}), but rather represents a centrally-extended
 version of this algebra. As a result, this new multiplet cannot be described
 in terms of the basic root superfields described above.
 The second sense in which dualization can
 be interpreted is more properly aligned with the duality maps
 described so far.  In this second sense we identify the designated
 fermionic node with the proper-time derivative of a ``dual" node.
 In this second sense, we obtain a multiplet
 which can be described by a root superfield, and which does represent
 the superalgebra (\ref{nextend}).  But this multiplet lies
 outside the root tree.  For reasons made more clear in the Appendix
 this multiplet requires a slight modification to
 the diagrammatics introduced to this point.  In cases where
 AD maps are implemented level-wise, the two senses of dualization
 described above coincide.  Otherwise, as we have just explained,
 theses senses differ, and each sense maps multiplets in the root
 tree into multiplets outside the root tree.

 First, lets consider dualization in the first sense.
 By reversing both arrows which connect to only one of the
 two level-one fermion nodes in the $\Omega_{2+}^{(2)}$
 we obtain the following new adinkra,
 \begin{center}
 \includegraphics[width=.75in]{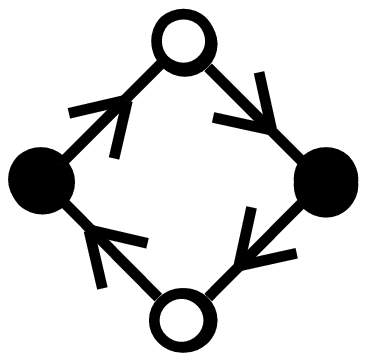}
 \end{center}
 This particular adinkra symbol has several noteworthy features.
 First of all, it is impossible to fold this adinkra into
 a linear form.  This is because of a topological obstruction
 which relates in an interesting way to the corresponding
 transformation rules.  As explained above, a given node
 is designated as ``lower" than another node if the second node
 can be reached from the first by using the arrows to define
 a flow pattern.  Nodes which are downstream in such a
 flow describe higher components.  For the interesting multiplet
 shown above, each node is at the same time both upstream and
 downstream of every other node; there is no highest component
 and there is no lowest component.
 We shall refer to multiplets with this feature as escheric
 multiplets, owing to the similarity with patterns found in
 many drawings of M.~C.~Escher.  Another, rather surprising
 feature is manifested by writing down the
 corresponding transformation rules using the procedure described
 above.  This is done in the Appendix.  It turns out
 that this multiplet does not strictly represent the $N=2$
 superalgebra (\ref{nextend}).  Instead, this represents a
 centrally-extended version of this superalgebra.  At the time
 of this writing, the implications of this are not perfectly clear.
 But we find this intriguing.

 Next, lets consider dualization in the second sense.  By
 writing one of the two level-one fermions in the
 $\Omega_{2+}^{(2)}$ adinkra as the proper-time derivative of a
 dual fermion, we obtain a new multiplet which does properly
 represent the $N=2$ superalgebra without a central charge.
 However, in this case the transformation rules for
 one of the fermions include the antiderivative
 of one of the boson fields.  The details are explained in the
 Appendix, where it is also shown that this multiplet includes
 as a sub-multiplet the $N=1$ root multiplet
 $(2,0)_+\cong (0,2)_+\cong (0,-1)_-\cong (-1,0)_-$.
 Since these labels include integers which are neither
 0 nor 1, this multiplet lies outside the $N=1$ root tree.
 Accordingly, this is not properly described by the adinkra
 symbols defined to this point.  (A relevant addendum to the
 notation introduced above is also included in the Appendix.)
 We refer to this sort of multiplet as a ``type II escheric
 multiplet" to distinguish these from the central charge
 escherics described previously.
 The appearance of the antiderivatives in the transformation rules
 is another circumstance which may have interesting relevance to
 physics, especially in cases where the corresponding
 field describes the coordinate on a compact dimension.

 In this paper we are concerned principally with the elements of
 the root tree.  As a result we will not describe escheric
 multiplets any further in this main text.  More details are
 included in the Appendix.  We intend to study these constructions
 further in ongoing work, and hope to have more to say on this topic in
 the future.

 \setcounter{equation}{0}
 \section{$N=3$ Adinkras}
 \label{n3ads}

~~~~ Consider the $N=3$ scalar multiplet described by (\ref{rules}).
 To be specific, choice the ${\cal GR}(4,3)$ matrices
 according to $L_1=R_1=i\,\s_1\oplus\s_2$,
 $L_2=R_2=i\,\s_2\otimes{\mathbb I}_2$ and
 $L_3=R_3=-i\,\s_3\otimes\s_2$.
 It is easy to translate the corresponding transformation rules
 into an adinkra symbol using the rules described above.  The
 result is
 \begin{center}
 \includegraphics[width=1in]{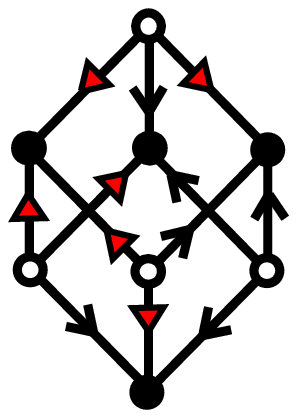} \,.
 \end{center}
 In this adinkra, the four bosonic and the four fermionic nodes are situated
 at the corners of a cube. Each of three quadruplets of parallel arrows corresponds to
 a different supersymmetry.

 In rendering the adinkra shown above we have distinguished the
 negative-parity arrows by giving these red color.  We have done this in
 order to make a couple of basic points. First of all, it is easy to verify
 the sum rule, described above, which says that the sum of the
 four arrow parities associated with any square sub-adinkra,
 must be odd.  This rule is easily verified on this
 diagram by tracing around each of the six faces of the cube,
 counting arrow parities in the process.  Next, recall that the
 arrow parities of every arrow connected to a given node flips
 when the node has its sign flipped.  In this way, the position of
 the negative parity arrows can be shifted around the adinkra
 symbol without changing the representation.  In the case of
 the cubic adinkra shown here, each time a node has its sign
 flipped, three arrows have their parity flipped.  This
 process flips either exactly zero or exactly two arrows in
 each subset of four arrows forming the edges of each face.  Since zero and
 two are even numbers this proves that the sum rule is maintained
 when any node has its sign flipped.

 Starting with the adinkra shown above, it is possible to cycle through
 a sequence of node flips, that cycles through all possible distributions
 of negative-parity arrows which satisfy the sum rule.  We will not describe
 a complete proof of this statement in this paper, however.  This shows
 that the ${\cal GR}(4,3)$ representation given above is unique.

 Consider next the $N=3$ base multiplet.  This has transformation
 rules given by (\ref{generic}).  If we translate these into an
 adinkra symbol we find that this symbol is identical to the $N=3$ scalar
 adinkra shown above.
 By following the folding rules described above, it is possible to
 reduce the $N=3$ base adinkra into a linear form.  This
 diagram can be folded as follows,
 \begin{center}
 \includegraphics[width=5.5in]{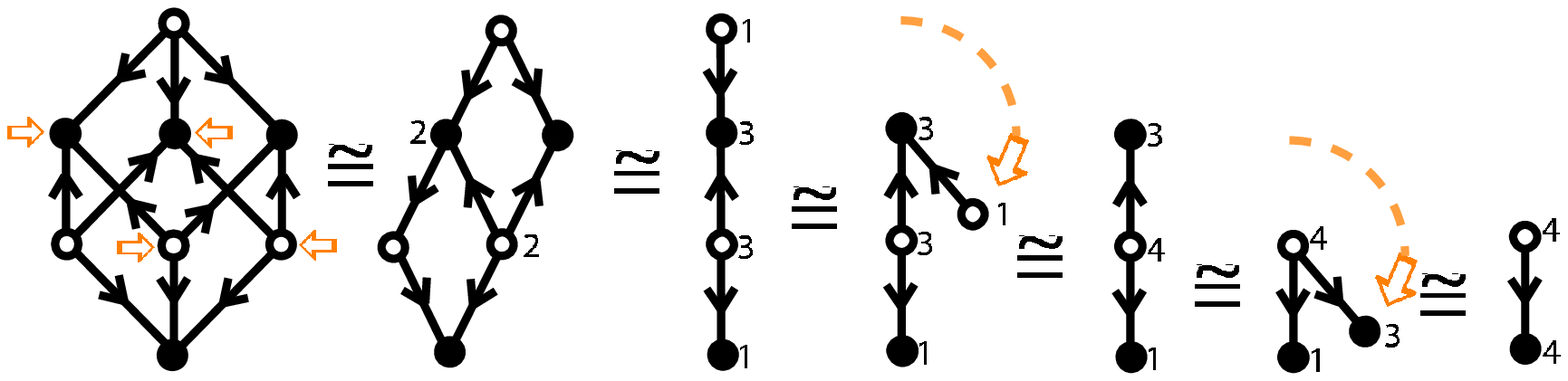}
 \end{center}
 In this sequence, we first pinch together two of the bosonic nodes
 and two of the fermionic nodes, as shown, thereby collapsing two
 opposite square faces into lines. This step reduces the figure to two
 dimensions.  Next we pinch the resultant multiplicity-two boson
 node together with another bosonic node, forming one
 multiplicity-three bosonic node.  At the same time we do
 a similar thing to form a multiplicity-three fermionic node.
 This reduces the adinkra into a linear form.  Two
 additional folds then transform the adinkra into
 its final form, given by two multiplicity-four
 nodes, one bosonic and one fermionic, connected by an arrow
 representing all three supersymmetries.

 The $N=3$ base adinkra has root label $(0000)_+$ and
 Omega designation $\Omega_{0+}^{(3)}$.
 The level-zero bosonic node corresponds to the topmost node
 appearing in the two-dimensional projection of the fully-unfolded
 form of adinkra shown above.
 Successive levels in the root superfield correspond
 to the sequence of horizontal node groupings which appear in
 this projection.
 We can form distinct multiplets starting with the $N=3$
 base adinkra, by performing AD maps and Klein flips.
 For example, we dualize at level-three by flipping
 all arrows which connect to the level-three boson
 (the bottommost node in the above diagram.)
 Doing this and then folding the resultant adinkra, we
 observe the following,
 \begin{center}
 \includegraphics[width=4.3in]{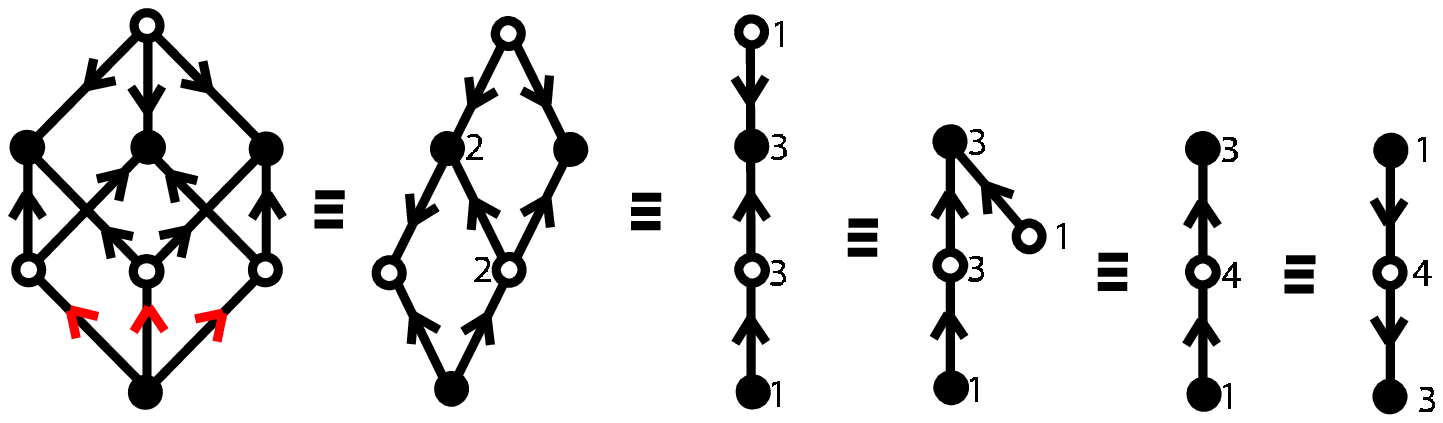}
 \end{center}
 In the final step we have rotated the adinkra by 180
 degrees, so that all arrows point downward.  In this case,
 we see that the arrow structure precludes the analog of the
 final fold made previously in the case of the $(0000)_+$
 adinkra.

 As another example, start with the $(0001)_+$ adinkra and dualize at
 level-two.  This generates the map $(0001)_+\to (0011)_+$.
 This is implemented by flipping
 all arrows which connect to the level-two nodes in the
 fully unfolded $(0001)_+$ adinkra.  In this way
 we obtain
 \begin{center}
 \includegraphics[width=2.5in]{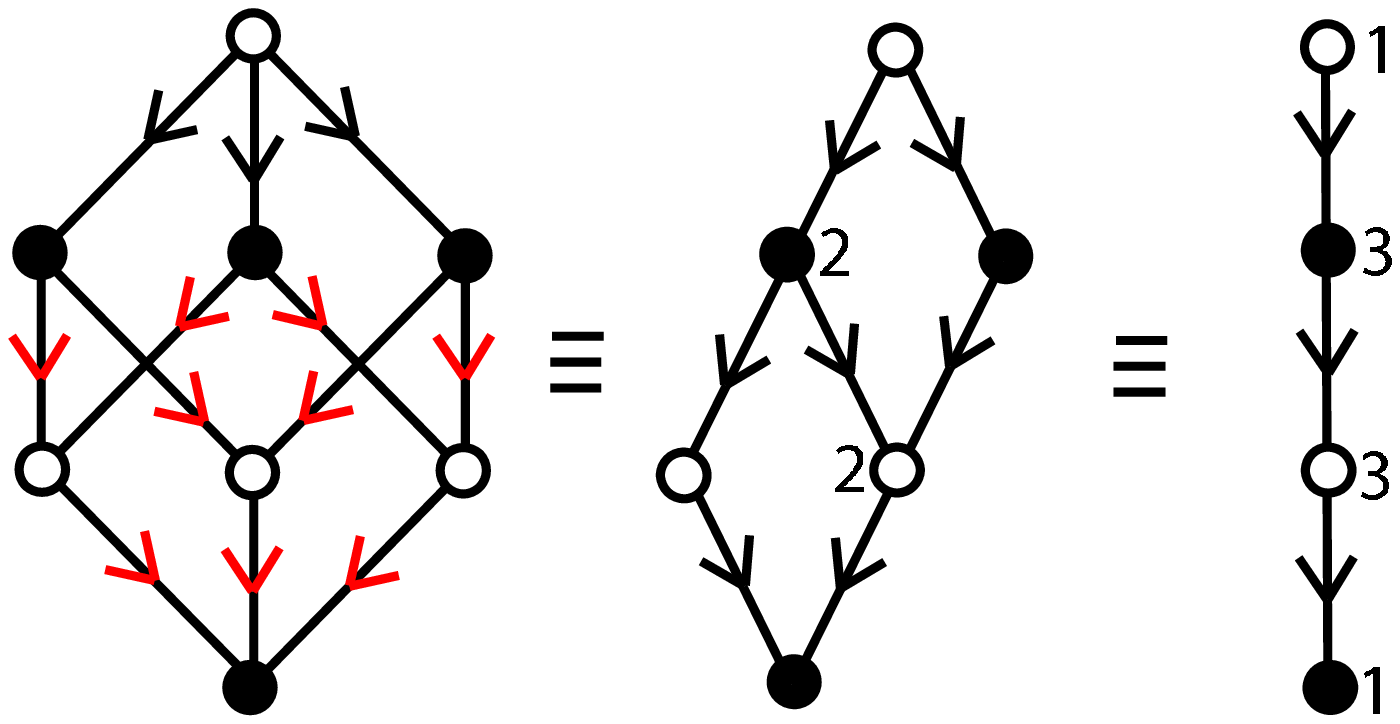}
 \end{center}
 Here fewer folds are permitted by the arrow
 structure than in the previous case, so that the fully-folded
 $(0001)_+$ adinkra has four nodes, rather than two.
 The maximal number of nodes in a fully folded adinkra is $N+1$.
 We denote the fully folded adinkra which has $N+1$ nodes as the
 ``top adinkra" for that value of $N$.

 As a final example, start with the $(0011)_+$ adinkra and dualize at
 level-zero.  This generates the map $(0011)_+\to (1011)_+\cong (0100)_+$.
 This is implemented by reversing all
 arrows connecting to the level-one nodes in the $(0011)_+$
 adinkra.  In this way we obtain
 \begin{center}
 \includegraphics[width=4.5in]{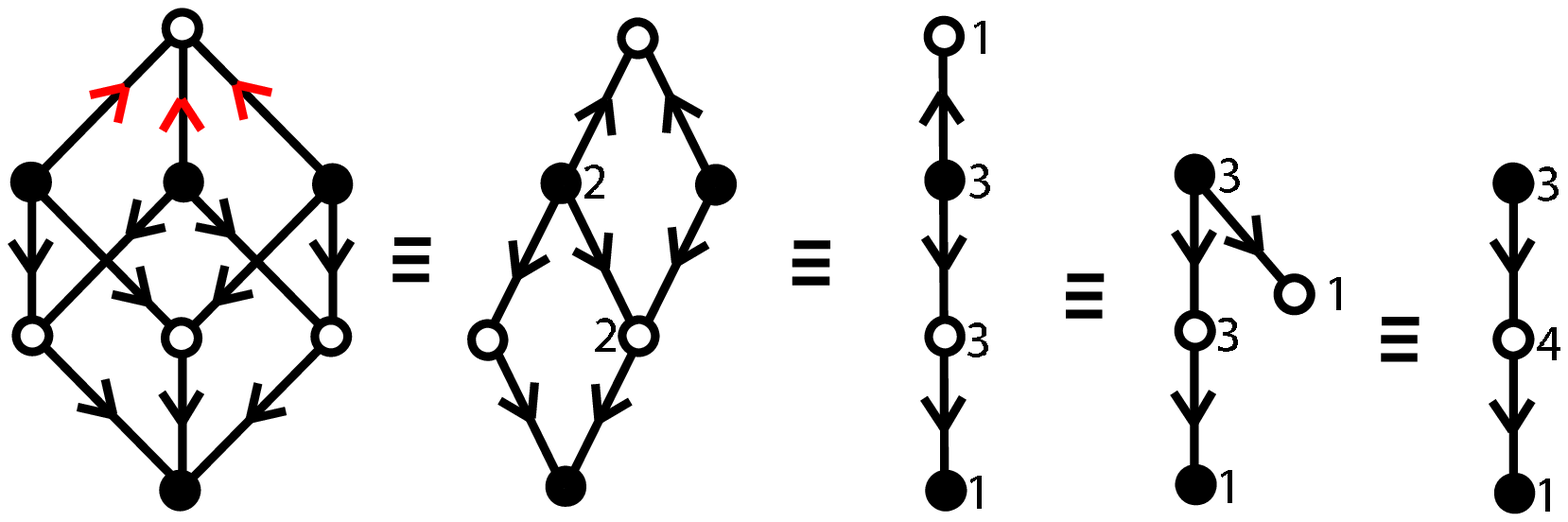}
 \end{center}
 In this case we see that the fully-folded adinkra has three
 compound nodes, having the multiplicities shown in the final
 diagram.

 As described previously, every adinkra in the root tree for any
 value of $N$ can be folded into a linear chain having $N+1$ nodes.
 Most adinkras can be folded further, so that in fully-folded form these
 exhibit fewer than $N+1$ compound nodes.  The number of compound nodes in
 the fully-folded form corresponds to the number of
 distinct component heights.  On the other hand, the partially-unfolded form
 describing a chain with $N+1$ nodes is more useful for
 implementing AD maps.  This is because in this form
 the adinkra nodes sequentially correspond to root superfield
 levels.  Since AD maps are implemented in a level-specific
 manner, these can be implemented on this form by reversing the
 compound arrows which connect to the nodes corresponding to
 desired levels.  To implement AD maps on elements of the root
 tree it is not necessary to unfold the diagram into more than
 one dimension.

 We define the ``depth" of an automorphism as the number of
 dimensions that an adinkra has to be unfolded into before the
 particular automorphism can be implemented.  Thus, the
 AD maps which we have described above
 comprise depth-zero automorphisms on the space of superalgebra
 representations. Arrow reversals and Klein flips at each depth
 greater than zero form separate abelian groups, since
 each of these operation generates its own ${\Z}_2$ subgroup.
 We define the ``rank" of a supermultiplet as one less than
 the minimal number of dimensions spanned by the fully folded adinkra. In this way the
 root tree comprises depth-zero multiplets.  The escheric
 multiplets described above correspond to multiplets having
 depth greater than zero.

 \setcounter{equation}{0}
 \section{The $N\le 3$ Root Trees and Auxiliary Fields}
 \begin{figure}
 \begin{center}
 \includegraphics[width=2in]{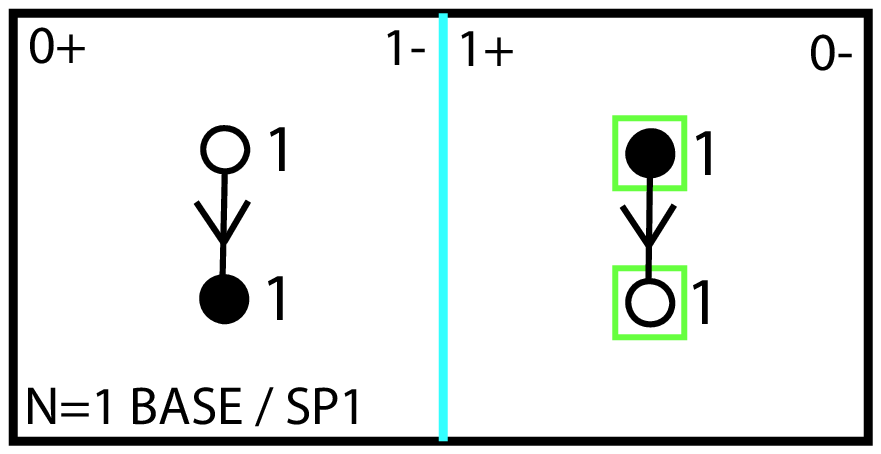}\\[.1in]
 \caption{The root-tree for the case $N=1$.}
 \label{scan1}
 \vspace{.2in}
 \includegraphics[width=2in]{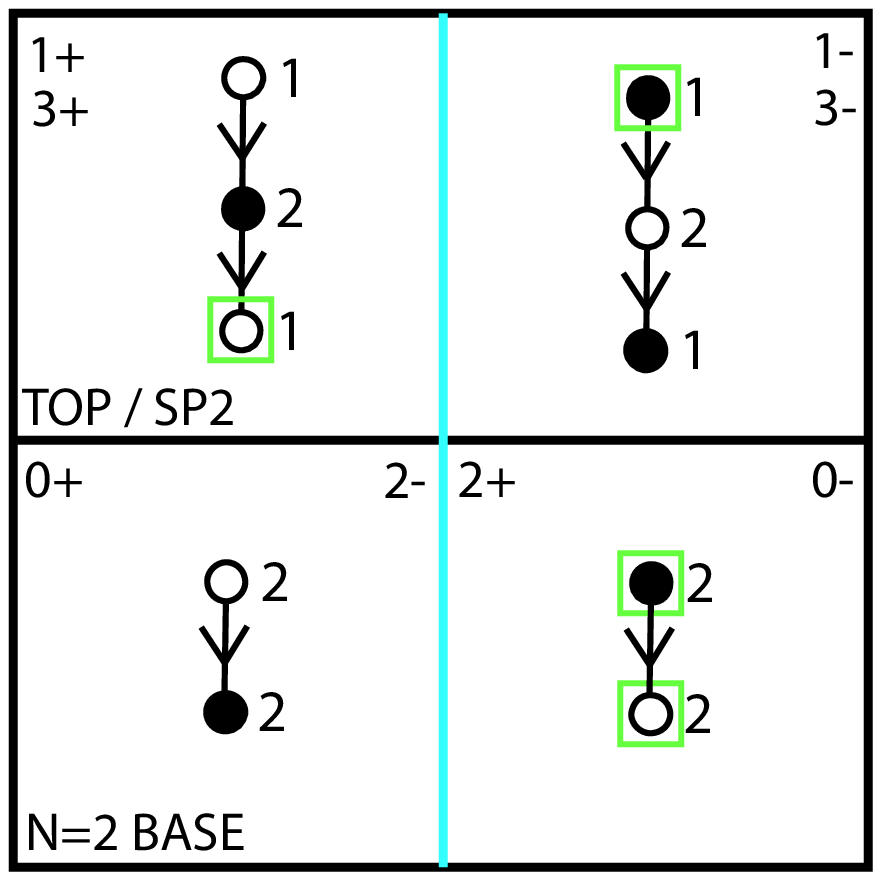}\\[.1in]
 \caption{The root-tree for the case $N=2$.}
 \label{scan2}
 \vspace{.2in}
 \includegraphics[width=4in]{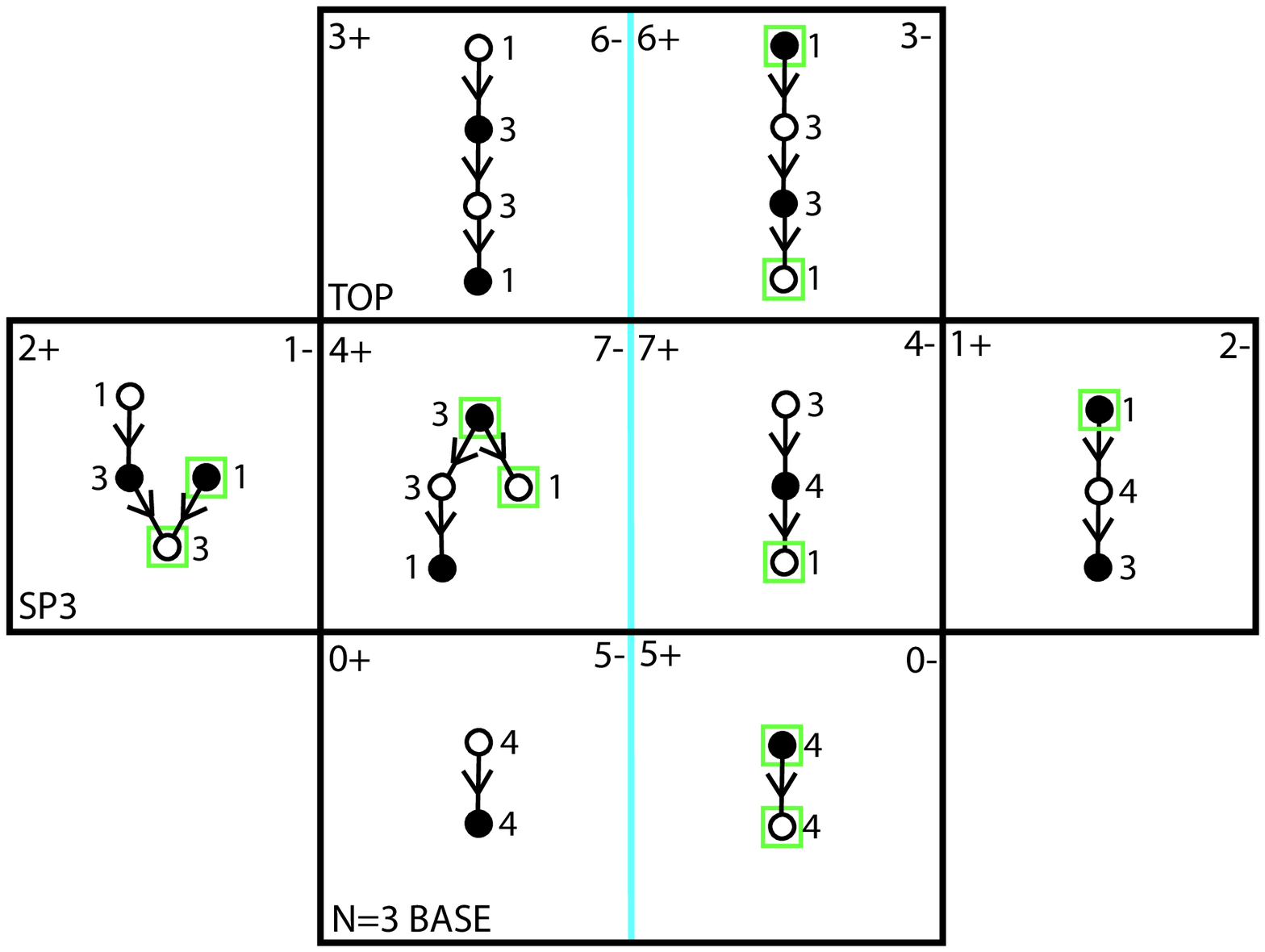}\\[.1in]
 \caption{The root-tree for the case $N=3$.}
 \label{scan3}
 \end{center}
 \end{figure}

~~~~ A field is typically deemed ``auxiliary" if it describes
 no dynamical (on-shell) degrees of freedom. However, it is possible to make
 an equivalent non-dynamical definition using
 the flow pattern generated by the arrows in adinkra symbols.
 According to this definition, auxiliary bosons are defined as
 bosonic nodes which appear as flow sinks, nodes to which all
 associated arrows point toward.  Auxiliary fermions are defined
 as fermionic nodes which appear as flow
 sources, nodes from which all associated arrows point away.
 For all supersymmetric actions with minimal kinetic derivatives,
 fields deemed auxiliary from the dynamics-free
 point of view are also auxiliary from the usual dynamical definition.
 As a notational convention, we sometimes
 place a box around auxiliary nodes in adinkra symbols.

 The 14 adinkras which describe the $N\le 3$ root trees are
 shown in Figures \ref{scan1}, \ref{scan2} and \ref{scan3}.
 These tables comprehensively exhibit the off-shell state counting
 for each of the minimal rank-zero multiplets, and also clearly
 indicate
 the interconnections between these generated by AD maps and Klein flips.
 For each choice of $N$, the adinkras are displayed in cells
 which include numbers describing the Omega designation for
 that multiplet.  For instance, the $N=3$ adinkra labelled $4+$ corresponds to
 $\Omega_{4+}^{(3)}$.  Extra numbers in any
 cell correspond to a notational redundancies, such as
 $\Omega_{7+}^{(3)}\cong\Omega_{4-}^{(3)}$.
 The root label for any of these multiplets are readily obtained by writing the decimal
 number in the Omega notation as the binary equivalent.  Accordingly, the root
 label for $\Omega_{4+}^{(3)}$ is $(0100)_+$.  This multiplet
 is obtained from the base multiplet $\Omega_{0+}^{(3)}$ by dualizing
 on the level-one nodes.   The reader is encouraged to verify the
 tables using the techniques described previously.

 Figures \ref{scan1}, \ref{scan2} and \ref{scan3} clearly exhibit the
 ${\Z}_2$ representation generated by Klein flips; adinkras on the
 right sides of each tabulation are obtained from those
 on the left by performing this operation.  The correspondence is
 easy to read, since the Klein flip
 manifests by flipping the sign on the Omega label.  The field multiplicity of each
 multiplet can be read off of the adinkras.  Bosons correspond to
 white nodes and fermions correspond to black nodes.  Boxed nodes
 correspond to auxiliary fields.  For instance, the multiplet
 $\Omega_{4+}^{(3)}$ is seen to have off-shell fields
 consisting of three physical bosons and one physical fermion
 and to have one auxiliary bosons and three auxiliary fermions.

 \setcounter{equation}{0}
 \section{$N=4$ Adinkras}

~~~~ New structures appear at $N=4$
 which are absent in the cases $N\le 3$.  The reason for this
 is that, in contrast to the cases $N\le 3$, the base
 multiplets $\Omega^{(4)}_{0+}$, and all representations obtained
 from this by AD maps and Klein flips, describe
 reducible representations.  For instance, the minimal
 $N=4$ multiplets have 4+4 off-shell degrees of freedom,
 as shown in Table \ref{dnvalues}.
 However, each element of the $N=4$ root tree has
 8+8 off-shell degrees of freedom, or twice the minimum.
 In this section we explain two methods for reducing such multiplets
 using the structure of adinkra symbols as a conceptual guide.  The first method
 uses consistent node identifications to describe the embedding of
 irreducible multiplets inside the root space.  The second method
 is to identify irreducible sub-adinkras describing gauge degrees-of-freedom.
 These methods prove sufficient for describing all known $N=4$ irreducible
 multiplets.

 The 18 adinkras which comprise the $N=4$ root tree are
 shown in Figure \ref{scan4}, which is structured in the same manner
 as Figures \ref{scan1}, \ref{scan2} and \ref{scan3}.
 The Omega designation for each multiplet is clearly indicated,
 including notational redundancies.
 \begin{figure}
 \begin{center}
 \includegraphics[width=6.5in]{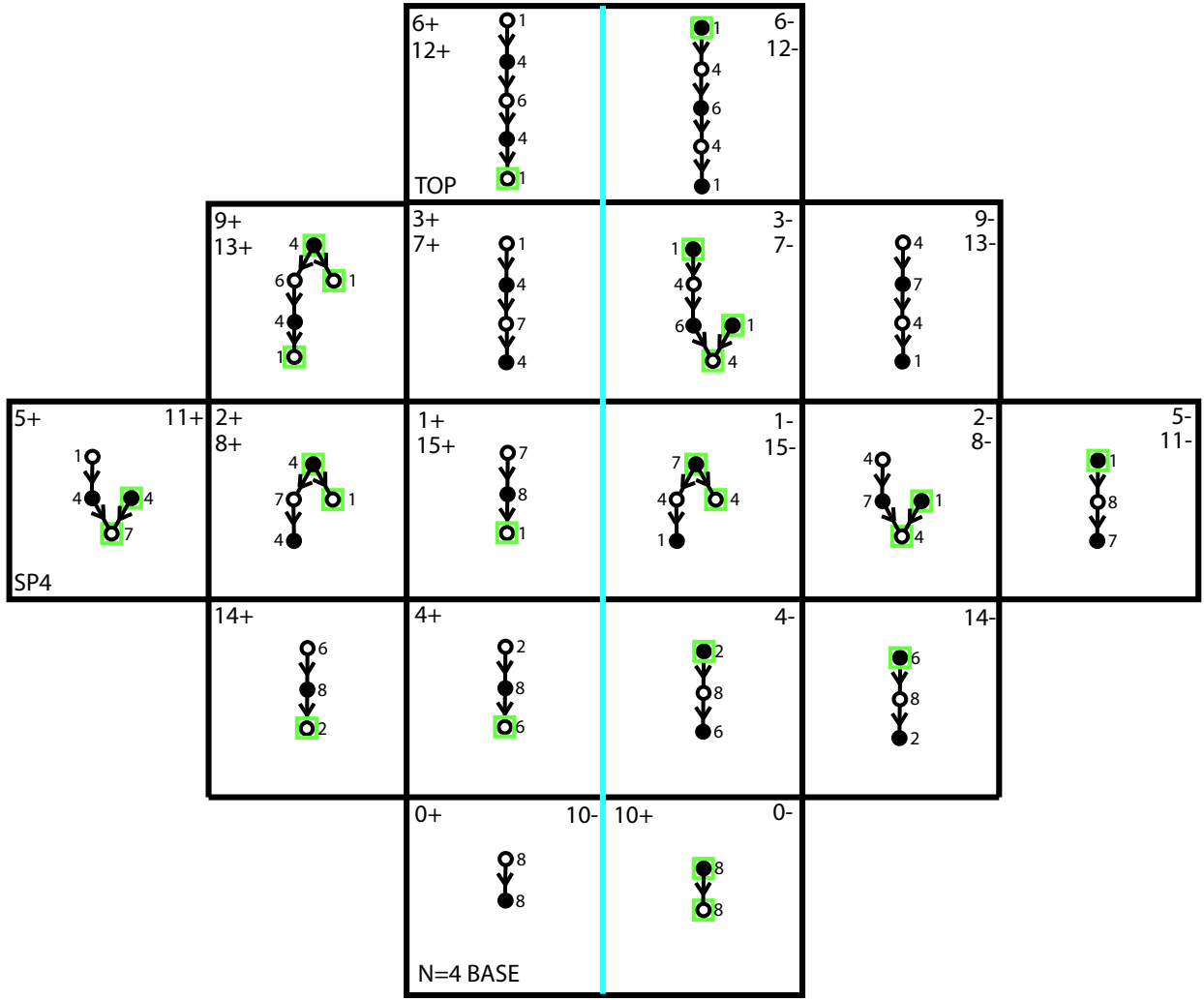}
 \caption{The $N=4$ root-tree.}
 \label{scan4}
 \end{center}
 \end{figure}
 Adinkras on the right side of Figure \ref{scan4} are obtained from those
 on the left by making a Klein flip.

 \subsection{Irreducible $N=4$ Multiplets}
~~~~ A class of irreducible multiplets is described by the scalar
 multiplets, with transformation rules given in (\ref{rules}).
 These are determined by choosing a representation of
 ${\cal GR}(d_N,N)$. For the case ${\cal GR}(4,4)$, we can make the choice
 $L_1=R_1=i\,\s_1\oplus\s_2$,
 $L_2=R_2=i\,\s_2\otimes{\mathbb I}_2$,
 $L_3=R_3=-i\,\s_3\otimes\s_2$ and
 $L_4=-R_4={\mathbb I}_2\otimes{\mathbb I}_2$.  It is
 possible to translate the transformation rules into
 an adinkra symbol, but there are extra subtleties not
 encountered in the cases $N\le 3$.

 The first subtlety is relatively simple.  Since each adinkra node
 connects with $N$-mutually-orthogonal arrows, one for each
 supersymmetry, it follows that the fully-unfolded form spans
 $N$-dimensions.  This makes the unfolded form relatively awkward
 to render on a page.  This problem can be overcome, at least for
 small values of $N$, by making small compromises with angles
 and with parallel lines, or it can overcome quite
 satisfactorily by folding the diagram down to
 one or two dimensions, if possible.

 The second subtlety has to do with combinatorics.
 For any value of $N$ the root multiplets have a total of
 $2^N$ nodes, just the right number to sit on the corners of
 an $N$-dimensional hypercube.  The irreducible multiplets
 have fewer than $2^N$ nodes, so it is not straightforward
 to connect the nodes with an $N$-dimensional orthogonal
 lattice. Many irreducible adinkras permit an embedding into
 an $N$-dimensional orthogonal lattice by including multiple
 copies of the original adinkra into the lattice.  Since there
 are 16 corners to a tesseract, and 8 total nodes in a scalar
 adinkra, it is conceivable that a double-copy of the scalar
 adinkra could fit properly into the tesseract.  In fact,
 this works perfectly well, as can be seen by the following
 diagram,
 \begin{center}
 \includegraphics[width=2.5in]{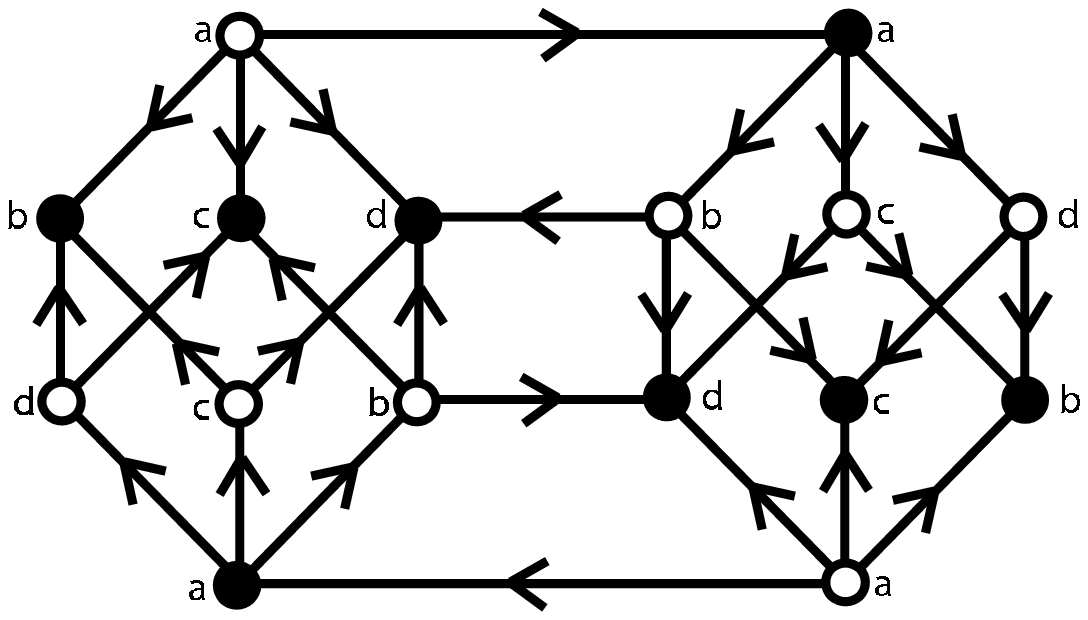} \,.
 \end{center}
 Here we have designated each node with a distinctive label.
 The cube on the right is an upside-down copy of the cube on the left.
 Thus the bosonic node $a$ on the top of the left side is the
 same as the bosonic node $a$ on the bottom of the right side.
 Each cube describes a representation of an $N=3$ subalgebra.
 The horizontal arrows describe the fourth supersymmetry.
 We have suppressed arrows pointing from the white $c$ node
 to the black $c$ node and from the white $d$ node to the
 black $d$ node, so as not to confuse the diagram.  This
 represents an accurate depiction of the transformation rules
 corresponding to the $N=4$ scalar multiplet described above.
 The adinkra diagram corresponding to the $N=4$ base multiplet is
 the same as the double-box diagram shown above, except with all
 of the identifications removed.  Thus, this method shows a way
 to embed the scalar multiplet into the $N=4$ root space.

 We can simplify the presentment of the scalar adinkra
 by folding the $N=3$ sub-diagram in the manner
 described in section \ref{n3ads}.  In this way, the diagram
 takes the simpler form
 \begin{center}
 \includegraphics[width=1.3in]{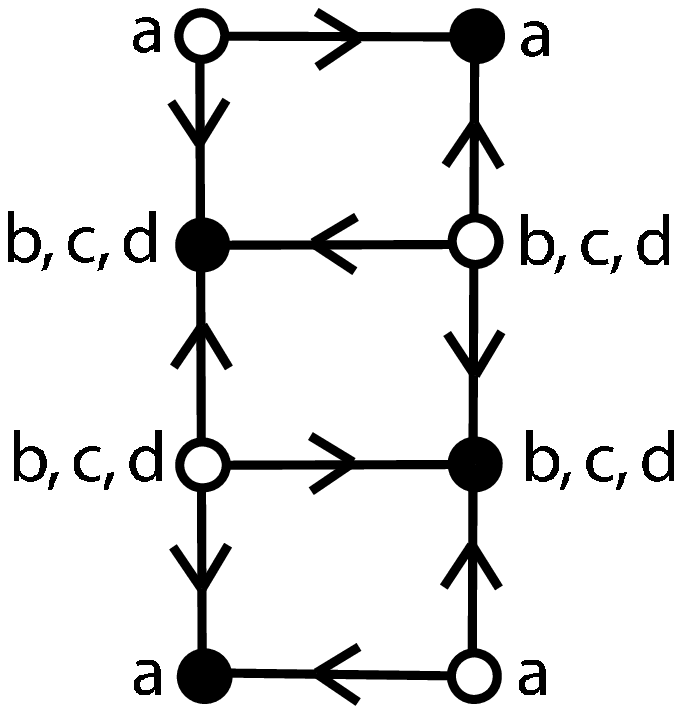}
 \end{center}
 In the folded form, the pairwise node identifications
 remain indicated by labels.  Another
 simplifying convention is to divide the two equivalent parts of
 this diagram using a ``mirror plane", and to
 redraw as follows,
 \begin{center}
 \includegraphics[width=1.2in]{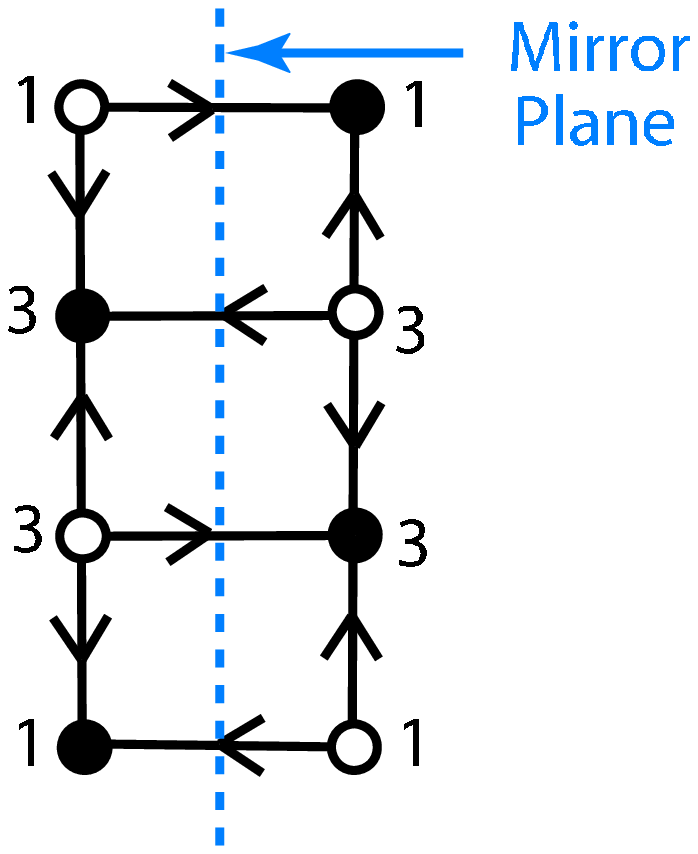}
 \end{center}
 Here we have replaced the labels $a,b,c$ and $d$ with the
 numerals representing node multiplicity.  The top node on the left side
 of the mirror is identified with the bottom node on the right side, the
 second node on the left is identified with the third node
 on the right, and so forth.  The mirror is inverting,
 since the object side is projected upside-down on the image
 side.

 In fact, there is some extra freedom
 in making these identifications: the image nodes can be identified with the
 object nodes with a change of sign.  There are four consistent
 ways to arrange this, denoted by including
 plus signs and minus signs on the mirror plane.
 Since the right side of the mirror is superfluous, this
 can be omitted when rendering the adinkra.
 The four possible multiplets obtained
 in this way have the following adinkras
 \begin{center}
 \includegraphics[width=3in]{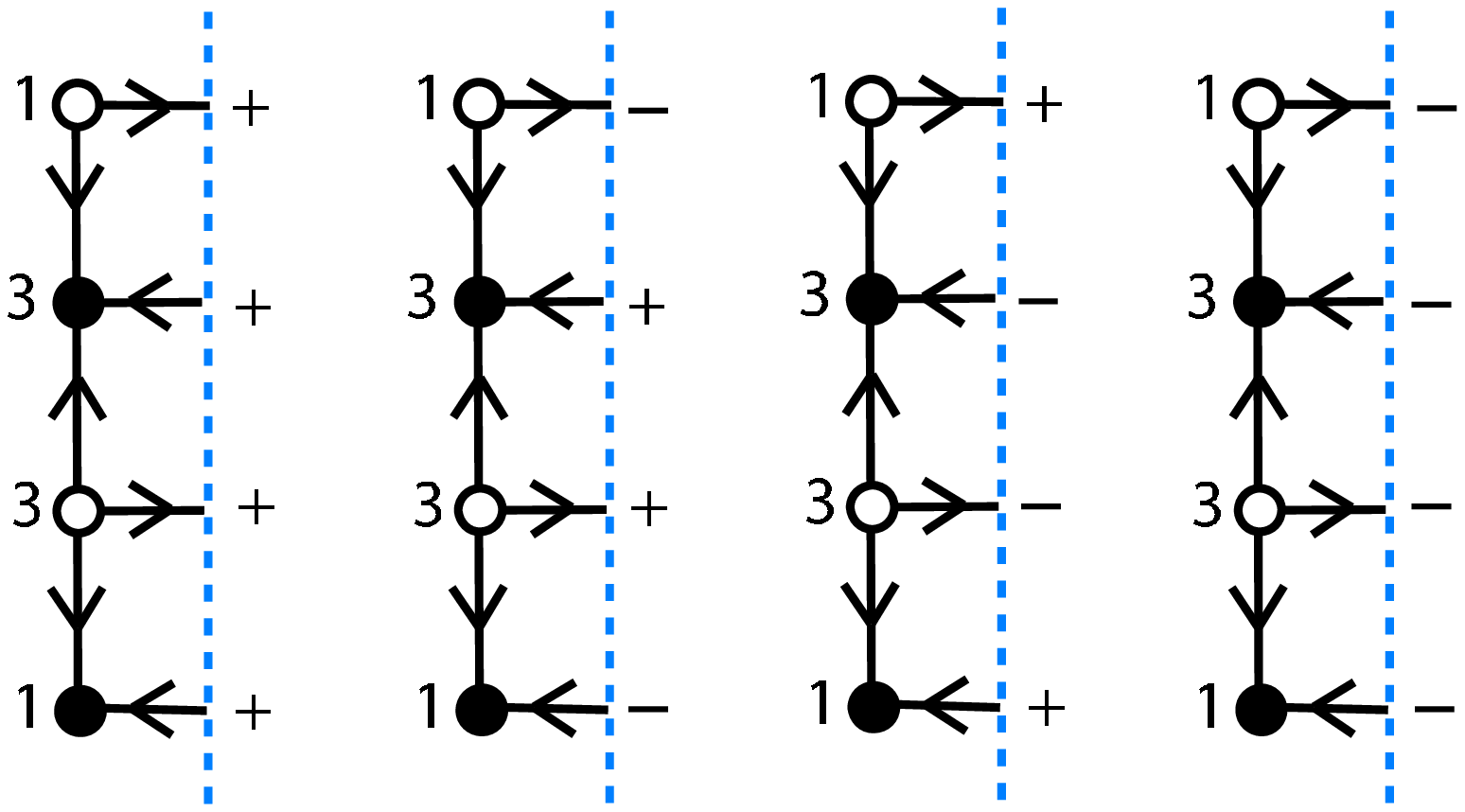}
 \end{center}
 The reason these are the only possibilities is that there
 is a consistency condition on the placement of the sign flips.
 Since the image is inverted, it follows that the bottom
 arrow is a continuation of the top arrow.  Similarly the
 two middle arrows are images of each other.  Thus, there are, in
 essence, only two independent choices of sign flips.
 The four multiplets obtained in this way describe
 the four separate scalar multiplets described
 in \cite{GatesKetov}. The different sign choices on the mirror plane
 describe different ways to assign arrow parity to the diagram.
 These four choices describe different elements of a particular
 conjugacy class of multiplets, a quaternionic analog of
 the difference between chiral and antichiral multiplets
 \footnote{Different elements of a given conjugacy
 class are often considered distinct.  For instance,\newline
 $~~~~~~$  chiral multiplets and antichiral multiplets in supersymmetric field
 theories describe two  \newline $~~~~~~$
 distinct elements of a common conjugacy class of
 representations.}.
 The first multiplet shown above
 is the same as the chiral multiplet described previously.
 It is possible to neglect arrow parity and use the undecorated
 adinkra symbols to describe multiplets as conjugacy classes.

 In fact, the manner in which we have organized the discussion of
 the $N=4$ scalar multiplets allows for a rigorous proof that the
 four scalar $N=4$ scalar multiplets are in fact distinct.
 The proof relies on the fact that once the $N=3$ sub-adinkras
 are given a particular arrow parity then every non-trivial inner
 automorphism of these representations alters this arrow parity
 assignment.  Inner automorphisms are generated by permutations
 of nodes and by sign flips.  Distinct multiplets are described
 by adinkras which cannot be mapped into each other by such
 operations. It is impossible to alter the parity of any arrow which crosses the
 mirror plane by virtue of node permutations or node sign flips
 and, at the same time, maintain the intrinsic arrow parity
 specific to the $N=3$ cube diagram on either side of the
 mirror plane.  We hope that this discussion may help alleviate
 skepticism regarding the multiplicity of $N=4$ scalar multiplets.

 There is another interesting way to fold the double-cube
 diagram shown above.  In this maneuver, we pinch the white
 $a$ node together with the white $c$ node, and at the same
 time pinch the black $a$ node together with the black $c$ node.
 The diagram then flattens into the following form
  \begin{center}
 \includegraphics[width=2.5in]{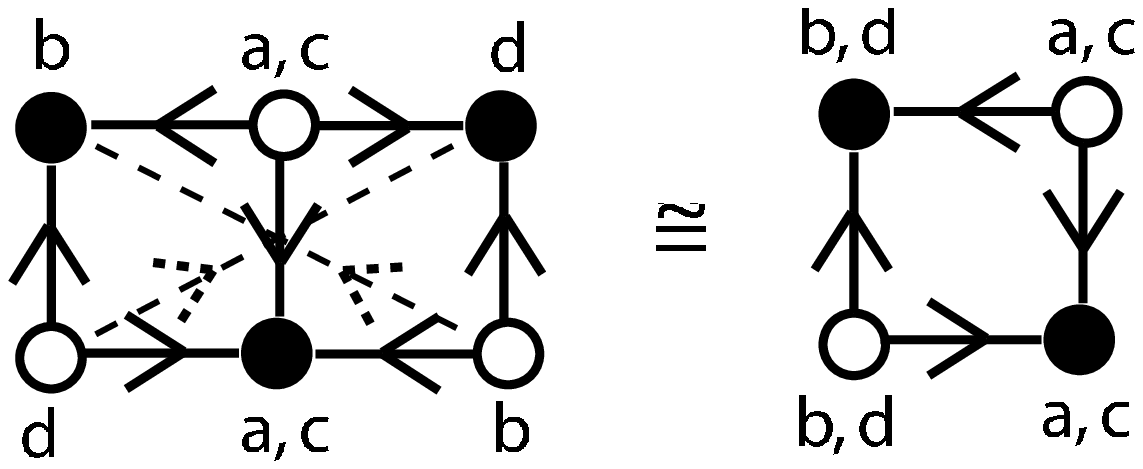} \,.
 \end{center}
 The dotted lines represent
 the fourth supersymmetry described by the horizontal lines in the
 double-cube diagram.  (Two more dotted lines are coincident with
 the vertical line connecting the $a$ and $c$ nodes.) We have
 folded this diagram a final time, by using the middle vertical
 line as a hinge, lifting the right vertical line out of the
 page and then placing this on top of the left vertical line.
 It is clear that the dotted lines land on top of each other with
 the proper orientation.  This is what enables this operation.
 We are left with a diagram which can be drawn as follows,
 \begin{center}
 \includegraphics[width=1.5in]{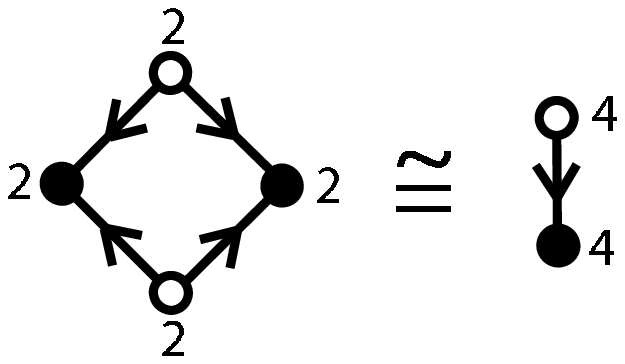} \,.
 \end{center}
 In this form, we see that this adinkra describes a pairwise
 assembly of the $N=2$ base adinkras, connected using parallel arrows
 representing a second pair of supersymmetries.
 This fully-folded form is more satisfactory
 for many purposes, as compared to the more complicated forms shown above.
 But the embedding inside of the root
 space given above is illuminating and useful in its own right.
 This multiplet corresponds to the shadow of the $d=1$ $N=4$
 linear multiplet.

 By drawing the analog of the double-cube adinkra for the
 root multiplet $\Omega_{1+}^{(2)}$ rather than for
 the base multiplet, and then going through a completely
 analogous sequence of identifications and folding steps,
 we arrive at an adinkra symbol which can be drawn as follows,
 \begin{center}
 \includegraphics[width=1.5in]{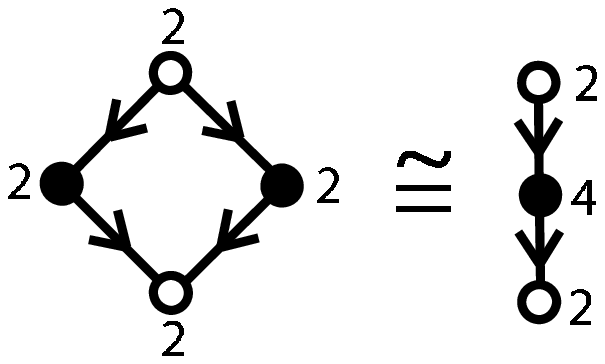} \,.
 \end{center}
 This adinkra is another example of a pairwise assembly of
 $N=2$ root multiplets.  This time it describes a pair
 of $\Omega_{1+}^{(2)}$ multiplets rather than a pair of
 $N=2$ base multiplets.
 This adinkra corresponds to the shadow of the
 $D=4$ $N=1$ chiral multiplet.
 Since we have suppressed arrow
 parity in this discussion, this more accurately
 actually describes the
 conjugacy class corresponding to the chiral multiplet.
 (Thus, this adinkra also depicts antichiral multiplet,
 if different choice of arrow parity is selected.)

 So far we have explained how the shadows of three of the four irreducible
 $D=4$ $N=1$ multiplets can be described using adinkra symbols.
 There is one more $D=4$ $N=1$ irreducible multiplet, however,
 the vector multiplet.  This is explained in the following
 subsection.

 \subsection{Gauge Invariance}
~~~~ Consider the reducible multiplet $\Omega_{6+}^{(4)}$, which is
 described by the top adinkra in the root tree.  This
 can be drawn in partially-folded and in folded form as
 follows,
 \begin{center}
 \includegraphics[width=2in]{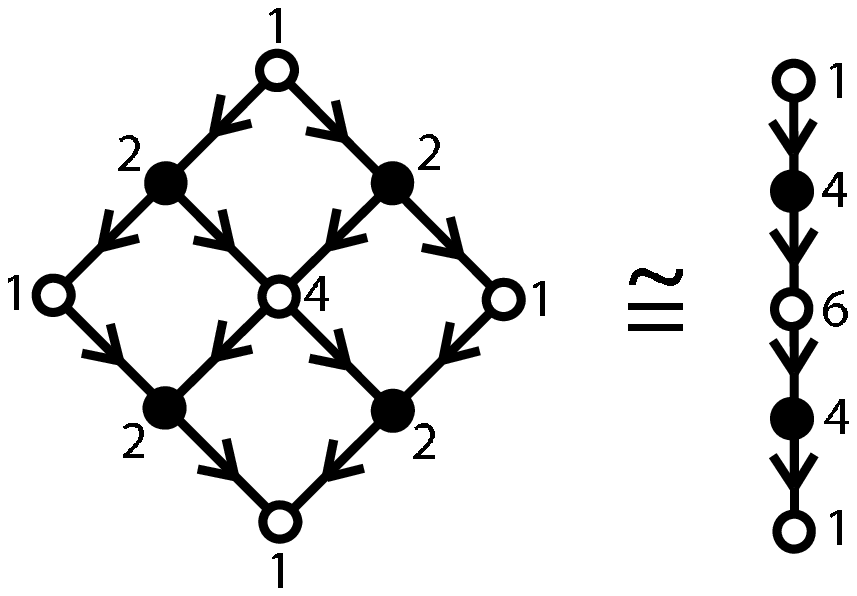} \,.
 \end{center}
 This adinkra exhibits yet another way to reduce degrees of
 freedom.

 A reducible adinkra can be re-defined by adding to it an irreducible adinkra,
 provided the structure of the smaller adinkra can be layered
 on top of the larger adinkra such that all arrows line up.
 In this way, one includes the larger multiplet into a class
 of multiplets related by a gauge transformation.  For example,
 we notice that chiral adinkra described previously
 fits onto the topmost diamond inside of the reducible
 $\Omega_{6+}^{(4)}$ adinkra.  We can represent the removal
 of the associated gauge degree of freedom using the following
 adinkra calculus,
 \begin{center}
 \includegraphics[width=5in]{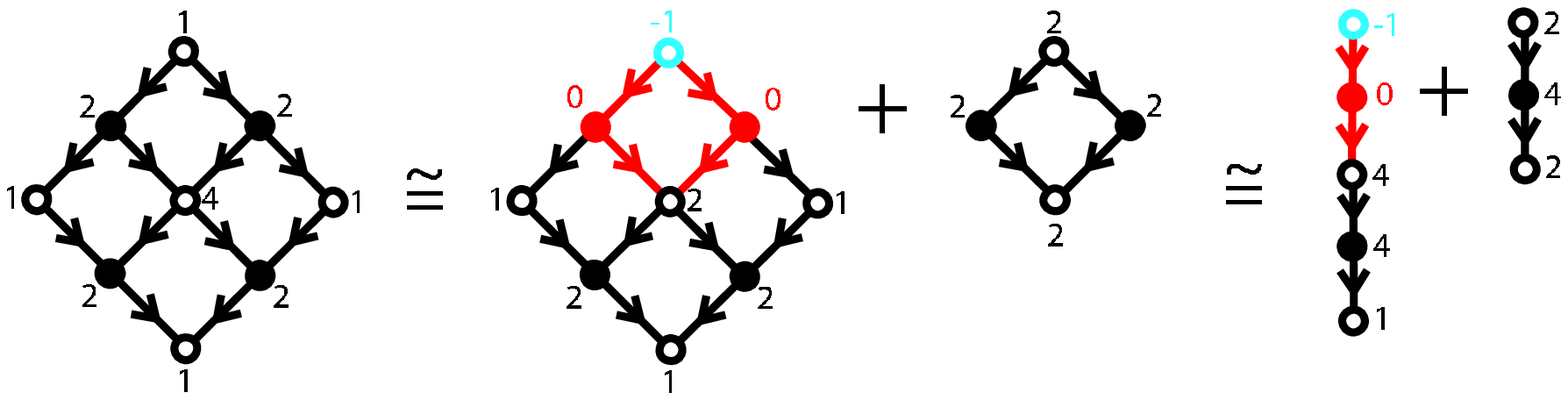} \,.
 \end{center}
 In this diagrammatic equation we see various noteworthy
 mnemonics at work.  First of all, we see how the structure of
 the chiral multiplet is embedded inside of the reducible
 vector multiplet.  Next, we see how node-by-node the degrees
 of freedom are subtracted.  Most noteworthy of all is the fact
 that the topmost node has been left with a formally negative
 field multiplicity.  What this means is that one of the two gauge
 degrees of freedom on the topmost node in the chiral multiplet has
 been used to
 remove the single degree of freedom at the topmost node in the
 vector multiplet.  The remaining degree of freedom in the chiral
 multiplet exists as a residual gauge degree of freedom after all
 of the possible node subtractions have been performed.  The
 residual gauge degree of freedom then ``flows" along the ghost
 structure as far as possible before finding itself on one of the
 un-removable nodes.  This node then exhibits the gauge degree of
 freedom in its multiplicity.  This demonstrates another rule for
 locating sub-adinkras which describe embedded gauge structures.
 Namely, the ``flow" of the gauge sub-adinkra must flow ``out" of
 the gauge sub-adinkra onto a non-removable node.  This example
 process is more concisely described
 in terms of fully-folded adinkras as follows,
 \begin{center}
 \includegraphics[width=3.5in]{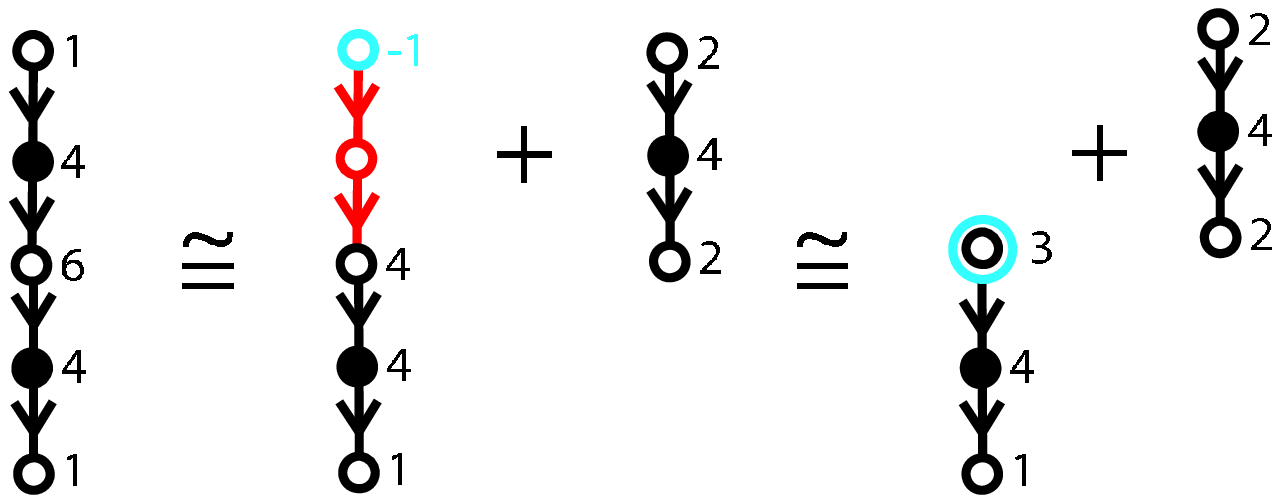} \,.
 \end{center}
 Here the extra circle on the gauge node indicates the
 ability to perform a gauge shift in the degree of freedom
 corresponding to this node.  This simple example is the
 diagrammatic representation of the shadow of the well-known
 Wess-Zumino gauge choice made in the context of $D=4$ $N=1$
 vector multiplets.

 \setcounter{equation}{0}
 \section{Spinning Particles}

~~~~ As a final example, we show how spinning
 particle multiplets can be described using adinkra symbols.
 Start with the base adinkra $\Omega^{(4)}_{0+}$, then dualize on
 the level-two and level-four bosons.  This produces the
 $\Omega^{(4)}_{5+}$ adinkra, given by
 \begin{center}
 \includegraphics[width=1.7in]{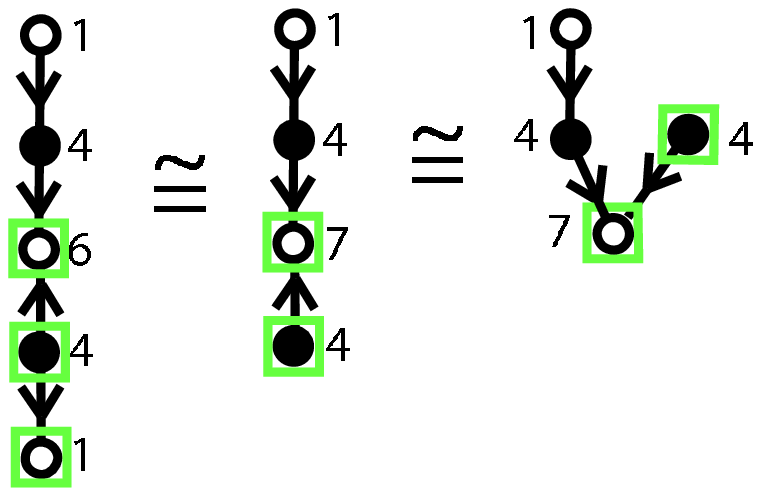}
 \end{center}
 Where in the final step we have oriented the nodes
 by height, so that all arrows point downward.
 As a rule, we keep the
 auxiliary fields separated in adinkra symbols.  This adinkra
 corresponds to the $N=4$ off-shell spinning particle
 multiplet first described in \cite{GatesRana1, GatesRana2}.

 By using a similar process, we can describe the ``Universal
 Spinning Particle Multiplet", also called the USPM, by drawing
 the base adinkra $\Omega_{0+}^{(N)}$, then dualizing on all
 bosons except at level-zero.  After a sequence of folds, this
 leads to the USPM adinkra,
 \begin{center}
 \includegraphics[width=1.8in]{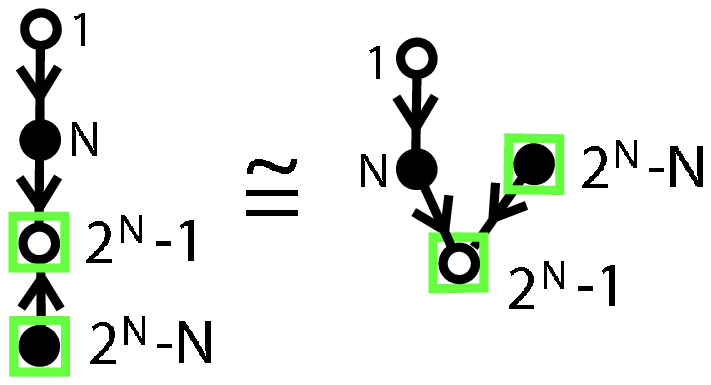} \,,
 \end{center}
 We see that the USPM has $2^N+2^N$ off-shell degrees of freedom,
 including $2N-1$ auxiliary bosons and $2^N-N$ auxiliary
 fermions.

 \setcounter{equation}{0}
 \section{Conclusions}

~~~~ We have described the rudiments of a symbolic method
 for organizing the representation theory of one-dimensional
 superalgebras.  This relies on special objects, which we
 have called adinkra symbols, which supply tangible geometric
 forms to the still-emerging mathematical basis underlying supersymmetry.
 We are optimistic that these symbols will prove useful in
 organizing a more rigorous and comprehensive representation
 theory for off-shell supersymmetry, not just in one-dimension but in
 higher dimensional field theories as well.

 As a demonstration of their power, we have used adinkras
 to codify, organize and reproduce all known minimal
 supermultiplets for the cases $N\le 4$.  Building on the concept
 of root superfields introduced in \cite{GatesLP}, we have
 used these symbols to interpret supersymmetry transformations
 in terms of flows on a corresponding root lattice.
 At the same time, we have shown how scalar multiplets and
 reduced chiral multiplets can be explained in terms of embeddings
 in these lattices, and have given an elegant description of
 gauge invariance in the case of a reduced Abelian vector
 multiplets.

 We have described a method for altering the appearance of
 adinkra symbols by folding.  This serves several purposes
 beyond the obvious one, to enable the rendering of
 multi-dimensional diagrams on a page.  Another use for
 folding adinkras is to allow a topological characterization
 of supermultiplets.  All of the known multiplets fit into the
 simplest category, described by diagrams which can be folded into
 a linear chain.  The existence of other multiplets, which we have
 called escheric multiplets, is curious, and we
 have to wonder what sorts of dynamics might be associated with these.

 The adinkra symbol for a given multiplet encodes the
 corresponding supersymmetry transformation rules.
 As a consequence, many cumbersome algebraic manipulations
 characteristic of supersymmetry calculations obtain a fresh
 look when phrased in terms of these symbols. We wonder if there
 might be a way to incorporate these adinkras so as to describe
 superfield dynamics as well.  Toward this end, we wonder how
 our technology should be modified to describe sigma models
 formulated on curved target spaces.

 It has been suggested that {\it all} supersymmetric theories in all
 dimensions are connected to each other by different sorts of dualities.
 The approach to one-dimensional supersymmetry centered on
 root superfield technology seem to substantiate this.
 From this point of view very many multiplets are interconnected
 by AD maps and by Klein flips.  This poses an intriguing
 dynamical riddle, however.  Using a chain of reasoning
 described in section \ref{admaps}, these duality maps correlate
 with sigma model target space dualities when an extra central
 term is switched on in the superalgebra.  The riddle is to obtain
  more comprehensive understanding of the relationships
 between automorphic dualities and geometric dualities, and
 to determine the role of supersymmetry central charges in this
 story. We feel that these observations are hinting at something
 fundamentally interesting.

 Future directions for this investigation include issues
 pertaining to supersymmetry representation theory and also
 issues pertaining to dynamics.  We intend to generalize the
 preliminary results described in this paper to include
 higher values of $N$, and to establish an understanding
 for how to ``oxidize" one-dimensional multiplets into
 higher-dimensional multiplets.  We would like to use this
 technology to study supergravity multiplets, and hopefully
 obtain an aesthetically pleasing alternative way to understand
 the ad-hoc superfield constraints which plague traditional
 approaches to this subject.  An unapologetic ambition which we
 have is to use this technology as a step towards finding an off-shell
 representation of $D=11$ supergravity.

 \vspace{.1in}
 \begin{center}
 \parbox{4in}{{\it ``Mathematics seems to endow one with something
 like a new sense.''}\,\,-\,\,Charles Darwin}
 \end{center}

 \vspace{.2in}

 \noindent
 {\bf Acknowledgements}\\[.1in]
 M.F is thankful to Donald Spector for stimulating discussions
 in an on-going collaboration
 regarding supersymmetric quantum mechanics, which sparked his
 original interest in this subject, and also to the Slovak
 Institute for Fundamental Research, Podvazie Slovakia, where
 much of this manuscript was prepared. S.J.G. wishes to acknowledge
 the staff and members of the Kvali Education Advising Center of the
 Tbilisi State University for assistance and hospitality during the period in
 this work began.

 \appendix

 \setcounter{equation}{0}
 \renewcommand{\thesection}{Appendix: Escheric Multiplets and Central Charges}
 \section{}
 \label{Appendix}
 \renewcommand{\thesection}{A}
 \renewcommand{\theequation}{A.\arabic{equation}}
 In this Appendix we provide a few details explaining some of the
 subtleties associated with the escheric multiplets described in
 section \ref{esmults}.  Although the main presentation of this paper
 concerns superalgebras without a central extension, we briefly
 indicate some connections with such extended superalgebras in
 this Appendix. Consider the centrally-extended $N=2$ superalgebra
 defined by
 \brr \{\,Q\,,\,Q^\dagger\,\}=H
      \hspace{.4in}
      Q^2=Z+i\,Y
      \hspace{.4in}
      [\,H\,,\,Z\,]=[\,H\,,\,Y\,]=0 \,,
 \label{algzy}\err
 where $H=i\,\der_\tau$ and $Z$ and $Y$ are Hermitian operators
 which comprise the real and imaginary parts of a
 supersymmetry central charge.  If we define a supersymmetry
 transformation via $\d_Q(\e)=\e\,Q+\e^\dagger\,Q^\dagger$,
 where $\e$ is a complex parameter, then
 (\ref{algzy}) can be re-written as
  \brr [\,\d_Q(\e_1)\,,\,\d_Q(\e_2)\,]=
      -2\,i\,\e_{[1}^\dagger\,\e_{2]}\,\der_t
      +2\,(\,\e_1\,\e_2+\e_1^\dagger\,\e_2^\dagger\,)\,Z
      -2\,i\,(\,\e_1\,\e_2-\e_1^\dagger\,\e_2^\dagger\,)\,Y \,.
 \err
 The real part of the central charge $Z$ appears in
 dimensionally-reduced field theories, where it
 appears as a shadow of internal momenta modes.
 The imaginary part of the central
 charge may have an algebraic connection with conformal supersymmetry.
 In \cite{Harmonic} implications of $Z\ne 0$ were studied,
 but $Y$ was constrained to vanish.

 In section \ref{esmults} we described two different sorts of
 duality operations which can be performed on only one of the
 two level-one nodes in the $\Omega_{2+}^{(2)}$ adinkra,
 and explained how each of these operations produce
 new topologically interesting multiplets.  In this following two
 subsections we present the transformation rules for these
 multiplets, in order to better substantiate the discussion in
 that section.

 \subsection{Type I Escherics}
 If we perform the first sort of duality transformation on the
 $\Omega_{2+}^{(2)}$ adinkra, by merely reversing the two arrows
 which connect with one of the two fermionic nodes, we obtain the following
 dual adinkra,
 \begin{center}
 \includegraphics[width=.8in]{Twisty.eps}
 \end{center}
 If we write down the transformation rules associated
 with this adinkra, by following the rules described in section
 \ref{n1sec}, we find that
 the algebra it represents includes a central extension.
 To see this, first determine the transformation rules from the
 diagram using the procedure described above,
 \brr \d_Q\,\phi_1 &=& i\,\e^1\,\dot{\psi}_1
      +i\,\e^2\,\psi_2
      \nonumber\\[.1in]
      \d_Q\,\psi_2 &=& \e^1\,\phi_2
      +\e^2\,\dot{\phi}_1
      \nonumber\\[.1in]
      \d_Q\,\phi_2 &=& i\,\e^1\,\dot{\psi}_2
      -i\,\e^2\,\psi_1
      \nonumber\\[.1in]
      \d_Q\,\psi_1 &=& \e^1\,\phi_1
      -\e^2\,\dot{\phi}_2 \,.
 \label{esch1}\err
 Here we have chosen one of the arrows to have negative parity in
 order to satisfy the proper sum rule for these parities.
 If we complexify the supersymmetry parameters by
 writing $\e=\e^1+i\,\e^2$ then the algebra satisfied on each of
 the component fields is the following
 \footnote{There is a difference between subscripts and superscripts on
 supersymmetry parameters: subscripts indicate different choices
 of parameters describing the same supersymmetry whereas superscripts
 distinguish supersymetries.}
 \brr [\,\d_Q(\e_1)\,,\,\d_Q(\e_2)\,]=
      -2\,i\,\e_{[1}^\dagger\,\e_{2]}\,\der_t
      -\ft12\,i\,(\,\e_1\,\e_2-\e_1^\dagger\,\e_2^\dagger\,)\,\d_Y
 \err
 where
 \brr \d_Y=(\,\der_t^2+1\,) \,.
 \err
 Notice that this multiplet includes a purely imaginary central
 charge, which acts in a non-trivial manner.  We refer to
 escheric multiplets with a non-trivial central charge as
 ``type I" escherics, to distinguish these from the different
 sorts of multiplets described below.

 \subsection{Type II Escherics}
 If we perform the second sort of duality transformation on the
 $\Omega_{2+}^{(2)}$ adinkra, by writing one of the fermionic nodes
 as the proper-time derivative of a dual fermion, we obtain
 transformation rules different than those described in
 (\ref{esch1}). In fact, in contrast to that type I escheric multiplet,
 the transformation rules obtained
 in this second way do obey the $N=2$
 superalgebra without a central charge.  To see this,
 start with the multiplet $\Omega_{2+}^{(2)}$
 and dualize on one of the two fermionic nodes
 by writing the corresponding fermion field as the proper-time
 derivative of a dual fermion, which we will now call
 $\psi_2$. This produces the following transformation rules,
 \brr \d_Q\,\psi_1 &=& -\e^1\,\phi_1
      +\e^2\,\int^t d\tilde{t}\,\phi_2(\tilde{t})
      \nonumber\\[.1in]
      \d_Q\,\phi_1 &=& -i\,\e^1\,\dot{\psi}_1
      +i\,\e^2\,\psi_2
      \nonumber\\[.1in]
      \d_Q\,\psi_2 &=& \e^1\,\phi_2
      +\e^2\,\dot{\phi}_1
      \nonumber\\[.1in]
      \d_Q\,\phi_2 &=& i\,\e^1\,\dot{\psi}_2
      +i\,\e^2\,\ddot{\psi}_1 \,.
 \label{es2a}\err
 It is easy to check that the algebra (\ref{algzy}) is satisfied
 with $Z=Y=0$ on each of the component fields.
 These rules can be described by the following adinkra symbol,
 \begin{center}
 \includegraphics[width=.8in]{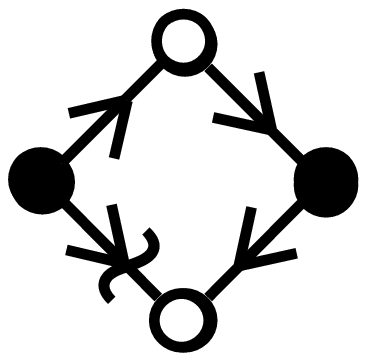}
 \end{center}
 where the newly distinctive type of arrow describes the
 multiplet $(2,0)_+\cong (0,2)_+\cong (0,-1)_-\cong (-1,0)_-$, a sort of
 $N=1$ multiplet not described in the main text.  The fact that the root
 labels for this multiplet include integers which are
 neither 0 nor 1 tell us that this multiplet is not
 in the root tree.
 This $N=1$ multiplet, which has the newly-defined adinkra
 \begin{center}
 \includegraphics[width=.2in]{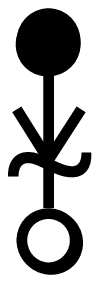}
 \end{center}
 has transformation rules
 \brr \d_Q\,\psi &=& \e\,\int^t d\tilde{t}\,\phi(\tilde{t})
      \nonumber\\[.1in]
      \d_Q\,\phi &=& i\,\e\,\ddot{\psi} \,.
 \label{newa}\err
 This describes an $N=1$ supermultiplet which is interestingly
 distinct from those in the $N=1$ root tree.  The presence of the
 antiderivatives in (\ref{es2a}) and (\ref{newa}) take on a
 particular topological significance if the boson which
 appears under these integrals describes a compact circular dimension.
 In this case, the antiderivative counts the number of times
 a particle winds around this circle.

\end{document}